REMARKS:
Below you find the *LaTeX source* of our recent paper.
The accompanying figures (in PostScript) are distributed separately in
an uuencoded compressed tar-file as required by the rules for
PostScript files on this bulletin board.

The LaTeX distribution is wrapped up in a single file with "--- snip ---"
lines delimiting the individual files. The corresponding file names are
indicated in the line ">>> <filename <<<" just above the leading "---
snip ---" lines.

The *full* distribution consists of the following files:

LaTeX part (this file):
-----------------------
2ladmqcd.tex:
   LaTeX source of the paper.

artcustom.sty, a4wide.sty, a4.sty:
   Required style files except epsf.sty. epsf.sty is of any use only in
   connection with the Rokicki dvips converter. a4wide.sty uses this
   particular version of a4.sty (with the later separately available at
   this bulletin board although I don't know whether the two are compatible).

PostScript part (distributed separately):
-----------------------------------------
fig1qcd.ps, fig2qcd.ps, fig3qcd.ps,
fig4qcd.ps, fig5qcd.ps, fig6qcd.ps, fig7qcd.ps:
   PostScript files read in by "2ladmqcd.tex" via epsf.sty provided with
   the Rokicki dvips DVI-to-PostScript converter. If you don't have the
   Rokicki dvips you should comment out the relevant epsf-commands in
   "2ladmqcd.tex".

Best regards,
             MARKUS LAUTENBACHER

MARKUS E. LAUTENBACHER,         | phone: +49/89/3209-2387
Technical University Munich,    | fax:   +49/89/32092296
Physics Dept.-Theo. Physics T31,| Internet:
D-8046 Garching, FRG.           | lauten@feynman.t30.physik.tu-muenchen.de

                           >>> 2ladmqcd.tex <<<
--- snip --- snip --- snip --- snip --- snip --- snip --- snip --- snip ---

\documentstyle[12pt,a4wide,artcustom,epsf]{article}

\newcommand{\dS}{\Delta S=1}
\newcommand{\dB}{\Delta B=1}
\newcommand{\dI}{\Delta I=1/2}

\newcommand{\ord}{{\cal O}}

\newcommand{\as}{\alpha_{\rm s}}
\newcommand{\aem}{\alpha}

\newcommand{\gf}{\gamma_5}

\newcommand{\tg}{\tilde{\gamma}}
\newcommand{\hg}{\hat{\gamma}}
\newcommand{\hG}{\hat{\Gamma}}

\newcommand{\gs}{\hat{\gamma}_{\rm s}^{(0)}}
\newcommand{\gem}{\hat{\gamma}_{\rm e}^{(0)}}

\newcommand{\gss}{\hat{\gamma}_{\rm s}^{(1)}}
\newcommand{\gse}{\hat{\gamma}_{\rm se}^{(1)}}

\newcommand{\gssndr}{\hat{\gamma}_{\rm s,NDR}^{(1)}}
\newcommand{\gsshv }{\hat{\gamma}_{\rm s,HV }^{(1)}}

\newcommand{\eps}{\varepsilon}
\newcommand{\epe}{\varepsilon'/\varepsilon}

\newcommand{\brackcc}[1]{\left[#1\right]_{\rm cc}}
\newcommand{\brackp }[1]{\left[#1\right]_{\rm p}}
\newcommand{\brackets}[2]{\left[ #1 \right]_{\rm #2}}

\newcommand{\rs}{\hat{r}_{\rm s}}

\newcommand{\drs}{\Delta\hat{r}_{\rm s}}

\newcommand{\rsndr}{\hat{r}_{\rm s,NDR}}

\newcommand{\rshv}{\hat{r}_{\rm s,HV}}

\newcommand{\Slash}{\not\!}

\newcommand{\tvs}{\vbox{\vskip 3mm}}
\newcommand{\svs}{\vbox{\vskip 5mm}}
\newcommand{\mvs}{\vbox{\vskip 8mm}}

\newcommand{\nn}{\nonumber}
\newcommand{\eqn}[1]{(\ref{#1})}

\newcommand{\newsection}[1]{\section{#1}\setcounter{equation}{0}}

\newcommand{\doctype}{}

\newcommand{\PreprintOrPaper}[2]{ \ifnum \doctype=1 #1 \else #2 \fi }

\begin{document}


\paperid{
{\sf MPI-PAE/PTh 106/92}\\
{\sf TUM-T31-18/92}
}

\date{\sf October 1992}

\author{\\
{\normalsize Andrzej J. BURAS${}^{1,2}$, Matthias JAMIN${}^{3}$,
 Markus E. LAUTENBACHER${}^{1}$, Peter H. WEISZ${}^{2}$}\\
\ \\
{\small\sl ${}^{1}$ Physik Department, Technische Universit\"at
M\"unchen,
                    D-8046 Garching, FRG.}\\
{\small\sl ${}^{2}$ Max-Planck-Institut f\"ur Physik
                    -- Werner-Heisenberg-Institut,}\\
{\small\sl P.O. Box 40 12 12, D-8000 M\"unchen, FRG.}\\
{\small\sl ${}^{3}$ Division TH, CERN, 1211 Geneva 23, Switzerland.}}

\title{
{\LARGE\sf
Two--Loop Anomalous Dimension Matrix for\\
$\dS$ Weak Non-Leptonic Decays {\rm I}:
${\cal O}(\as^{2})$}\footnote{Supported by the German
Bundesministerium f\"ur Forschung und Technologie under contract 06 TM 761
and by the CEC Science project SC1-CT91-0729.}
}

\maketitle

\begin{abstract}
\noindent
We calculate the two--loop $10 \times 10$ anomalous dimension matrix
${\cal O}(\as^{2})$ involving current--current operators, QCD
penguin operators, and electroweak penguin operators especially
relevant for $\dS$ weak non--leptonic decays, but also important for
$\dB$ decays. The calculation is performed in two schemes for
$\gamma_{5}$:  the dimensional regularization scheme with anticommuting
$\gamma_{5}$ (NDR), and in the 't Hooft--Veltman scheme.  We
demonstrate how a direct calculation of diagrams involving $\gf$ in closed
fermion loops can be avoided thus allowing a consistent calculation in
the NDR scheme.  The compatibility of the results obtained in the two
schemes considered is verified and the properties of the resulting
matrices are discussed. The two--loop corrections are found to be
substantial. The two--loop anomalous dimension matrix ${\cal
O}(\aem\as)$, required for a consistent inclusion of
electroweak penguin operators, is presented in a subsequent
publication.
\end{abstract}

\newpage
\setcounter{page}{1}


\newsection{Introduction}

Effective low energy Hamiltonians for non--leptonic weak decays
of hadrons are usually written as linear combinations of four--quark
operators. The coefficients of these operators, the
Wilson coefficient functions, can be calculated in the
renormalization group improved perturbation theory, as long as
the normalization scale $\mu$ is not too low. The size of these
coefficients depends on $\mu$ through the QCD effective coupling
constant and on the anomalous dimensions of the operators in
question. Since the operators generally mix under renormalization,
one deals with anomalous dimension matrices rather than with single
anomalous dimensions.

In the case of the $\dS$ Hamiltonian, relevant for instance for the $\dI$
rule and the ratio $\epe$, there are ten operators $Q_{i}, \;
i=1,\ldots,10$  which have to be considered. Consequently one deals
with a $10\times 10$ anomalous dimension matrix.
The ten operators in question can be divided into three classes:
\begin{itemize}
\item
current--current operators $Q_{i}, \; i=1,2$ originating in the usual
W--exchange and subsequent QCD corrections,
\item
QCD penguin operators $Q_{i}, \; i=3,\ldots,6$ originating in
QCD penguin diagrams, and
\item
electroweak penguin operators $Q_{i}, \; i=7,\ldots,10$ originating
in the photon penguin diagrams.
\end{itemize} \noindent
Explicit expressions for these operators are given in eq. \eqn{eq:2.1}.

Because of the presence of electroweak penguin operators a
consistent analysis must involve anomalous dimensions resulting
from both strong and electromagnetic interactions. Working to
first order in $\aem$ but to all orders in $\as$ the following anomalous
dimension matrix is needed for the leading and next--to--leading
logarithmic approximation for the Wilson coefficient functions,
\begin{equation}
\hg \; = \; \frac{\as}{4\pi} \, \gs +
\frac{\aem}{4\pi} \, \gem +
\frac{\as^{2}}{(4\pi)^{2}} \, \gss +
\frac{\aem\as}{(4\pi)^{2}} \, \gse \;.
\label{eq:1.1}
\end{equation}

In the leading logarithmic approximation only the one--loop matrices
$\gs$ and $\gem$ have to be considered. Inclusion
of next--to--leading corrections requires the evaluation of the two--loop
matrices $\gss$ and $\gse$. In addition some
one--loop finite terms are needed.

What is known in the literature are the matrices $\gs$ and
$\gem$
\cite{gaillard:74}--\nocite{altarelli:74,vainshtein:77,gilman:79,guberina:80,bijnenswise:84,burasgerard:87,sharpe:87,lusignoli:89,flynn:89}\cite{buchallaetal:90}
and a $2 \times 2$ submatrix of $\gss$ involving the current--current
operators $ Q_{1}$ and $Q_{2}$ \cite{altarelli:81,burasweisz:90}. The
purpose of the present and a subsequent paper is to complete the
evaluation of the matrix $\gss$ and to calculate $\gse$. Together with
certain one--loop finite terms this will allow the extension of the
phenomenology of non--leptonic $\dS$ transitions beyond leading
logarithms. The results for the $6 \times 6$  submatrix of $\gss$
involving current--current and QCD penguin operators together with some
phenomenological implications have already been presented by us in ref.
\cite{burasetal:92a}. In the present paper we give the details of these
two--loop calculations and generalize them to the full matrix $\gss$.
The matrix $\gse$ is considered in a subsequent publication
\cite{burasetal:92c}. In particular, we present the results for an
arbitrary number of colours, $N$, which is useful for the applications
of the $1/N$ expansion. The phenomenological implications of these new
results will be discussed in ref.~\cite{burasetal:92d}.

The calculation of the matrices $\gss$ and $\gse$
is very tedious. It involves a large number of two--loop diagrams which
make the use of an algebraic computer program almost mandatory
\footnote{The algebraic computer program for the $\gamma$-algebra TRACER,
written in MATHEMATICA \cite{wolfram:91}, which we extensively used during
the calculation, has been developed by two of us \cite{jamin:91} and is
available from the authors.}.
Moreover one has to deal with several subtleties which are absent in
the calculations of $\gs$ and $\gem$. First of all,
the two--loop anomalous dimensions depend on the renormalization scheme
for operators and the treatment of $\gf$ in $D \not= 4$ space--time
dimensions. It is known
for instance on the basis of ref. \cite{burasweisz:90},  that the $2
\times 2$  submatrix of $\gss$ involving $Q_{1}$ and $Q_{2}$
calculated in the NDR scheme (naive dimensional regularization with
anticommuting $\gf$) differs substantially from the one obtained in
the HV scheme (non anticommuting $\gf$ \cite{thooft:72,breitenlohner:77}).
The same feature is found for the full matrices $\gss$ and
$\gse$. The important point is that the difference between
$\hg^{(1)}$ calculated in two different schemes is on
general grounds entirely given in terms of $\hg^{(0)}$ and the
finite parts of one--loop diagrams calculated in the two schemes in
question. This relation is very useful for checking the compatibility
of two--loop calculations performed in different renormalization
schemes and plays an important role in demonstrating the scheme
independence of physical quantities. It is given in eq. \eqn{eq:2.12}.
Another subtle point is the dependence of the two--loop anomalous
dimensions obtained in the NDR scheme on whether a given operator has
been put in colour singlet or colour non--singlet form such as
\begin{equation}
Q_{2} \; = \; \left( \bar s_{\alpha} u_{\alpha} \right)_{\rm V-A}
              \left( \bar u_{\beta}  d_{\beta}  \right)_{\rm V-A}
\qquad {\rm or} \qquad
\widetilde{Q}_{2} \; = \; \left(\bar s_{\alpha} d_{\beta} \right)_{\rm V-A}
                          \left(\bar u_{\beta}  u_{\alpha}\right)_{\rm V-A}
\; .\label{eq:1.2}
\end{equation}
In $D=4$ dimensions $Q_{2}$ and $\widetilde{Q}_{2}$ would be equivalent by
means of Fierz transformation. Their one--loop anomalous dimensions
are the same.

Again the two--loop anomalous dimensions for these two forms of
operators can on general grounds be related to each other by
calculating one--loop diagrams. This feature turns out to be fortunate
in that it offers one way to deal with dangerous closed fermion loops
involving odd numbers of $\gf$ in two--loop diagrams. The latter are
known to be ambiguous in the case of the NDR scheme. Indeed, as we will
demonstrate explicitly in section~5 and in ref.~\cite{burasetal:92c}, it
is possible by using different forms of the operators and the relation
mentioned above, to express all diagrams with closed fermion loops in
terms of diagrams not containing such loops. In this way, consistent
calculations of $\gss$ and $\gse$ can be performed
in the NDR scheme and compared with the one obtained in the HV scheme,
free of $\gf$ problems.

The following sections constitute the calculation of $\gss$
and a detailed discussion of the subtle points mentioned above. In ref.
\cite{burasetal:92c}, the methods developed here are generalized to
include photon exchanges and are used to calculate $\gse$.

Our paper is organized as follows: In section~2, we give a list of
the ten operators in question, and we classify the one and two--loop
diagrams into current--current and penguin diagrams. We also discuss
the basic formalism necessary for the calculation of $\gss$.
In section~3, we recall the matrix $\gs$, and we calculate the
finite terms of one--loop diagrams in the NDR and HV scheme. In
section~4, the calculations and results for two--loop current--current
diagrams are presented. In section~5, an analogous presentation is
given for two--loop penguin diagrams. In section~6, we combine the
results of the previous sections to obtain $\gss$ in NDR and HV
schemes. We discuss various properties of this matrix,
in particular its Large-$N$ limit. Section 7 contains a brief summary
of our paper. In appendices A and B explicit expressions for the
elements of the $10 \times 10$ matrices $\gs$ and $\gss$ for an
arbitrary number of colours ($N$) and flavours ($f$) are given.
In appendix C the corresponding results for $\gss$ in the case of $N=3$
are presented.

\newsection{General Formalism}

\subsection{Operators}

The ten operators considered in this paper are given as follows
\begin{eqnarray}
Q_{1} & = & \left( \bar s_{\alpha} u_{\beta}  \right)_{\rm V-A}
            \left( \bar u_{\beta}  d_{\alpha} \right)_{\rm V-A}
\, , \nn \\
Q_{2} & = & \left( \bar s u \right)_{\rm V-A}
            \left( \bar u d \right)_{\rm V-A}
\, , \nn \\
Q_{3} & = & \left( \bar s d \right)_{\rm V-A}
   \sum_{q} \left( \bar q q \right)_{\rm V-A}
\, , \nn \\
Q_{4} & = & \left( \bar s_{\alpha} d_{\beta}  \right)_{\rm V-A}
   \sum_{q} \left( \bar q_{\beta}  q_{\alpha} \right)_{\rm V-A}
\, , \nn \\
Q_{5} & = & \left( \bar s d \right)_{\rm V-A}
   \sum_{q} \left( \bar q q \right)_{\rm V+A}
\, , \nn \\
Q_{6} & = & \left( \bar s_{\alpha} d_{\beta}  \right)_{\rm V-A}
   \sum_{q} \left( \bar q_{\beta}  q_{\alpha} \right)_{\rm V+A}
\, , \label{eq:2.1} \\
Q_{7} & = & \frac{3}{2} \left( \bar s d \right)_{\rm V-A}
         \sum_{q} e_{q} \left( \bar q q \right)_{\rm V+A}
\, , \nn \\
Q_{8} & = & \frac{3}{2} \left( \bar s_{\alpha} d_{\beta} \right)_{\rm V-A}
         \sum_{q} e_{q} \left( \bar q_{\beta}  q_{\alpha}\right)_{\rm V+A}
\, , \nn \\
Q_{9} & = & \frac{3}{2} \left( \bar s d \right)_{\rm V-A}
         \sum_{q} e_{q} \left( \bar q q \right)_{\rm V-A}
\, , \nn \\
Q_{10}& = & \frac{3}{2} \left( \bar s_{\alpha} d_{\beta} \right)_{\rm V-A}
         \sum_{q} e_{q} \left( \bar q_{\beta}  q_{\alpha}\right)_{\rm V-A}
\, , \nn
\end{eqnarray}
where $\alpha$, $\beta$ denote colour indices ($\alpha,\beta=1,\ldots,N$)
and $e_{q}$ are quark charges. We have omitted colour indices for the
colour singlet operators. $(V\pm A)$ refers to $\gamma_{\mu}
(1\pm\gf)$.  This basis closes  under QCD and QED renormalization and
is complete if external momenta and masses are neglected. However, at
intermediate stages of the calculation, we have to retain operators
which vanish on--shell. This will be discussed in detail in
section~5.2.

At one place, it will become useful to study a second basis in which the
first two operators are replaced by their Fierz conjugates,
\begin{eqnarray}
\widetilde{Q}_{1} & = &
\left( \bar s d \right)_{\rm V-A}
\left( \bar u u \right)_{\rm V-A}
\, , \nn \\
\widetilde{Q}_{2} & = &
\left( \bar s_{\alpha} d_{\beta}  \right)_{\rm V-A}
\left( \bar u_{\beta}  u_{\alpha} \right)_{\rm V-A}
\, ,
\label{eq:2.2}
\end{eqnarray}
with the remaining operators unchanged. In fact, the latter basis is
the one used by Gilman and Wise \cite{gilman:79}. We prefer however to
put $Q_{2}$ in the colour singlet form as in eq. \eqn{eq:2.1}, because
it is this form in which the operator $Q_{2}$ enters the tree level
Hamiltonian. Let us finally recall that the Fierz conjugates of the
$(V-A)\otimes (V+A)$ operators $Q_{6}$ and $Q_{8}$ are given by
\begin{eqnarray}
\widetilde{Q}_{6} & = & -\,8
\sum_{q} \left( \bar s_{\rm L} q_{\rm R} \right)
         \left( \bar q_{\rm R} d_{\rm L} \right)
\, , \nn \\
\widetilde{Q}_{8} & = & -\,12
\sum_{q} e_{q} \left( \bar s_{\rm L} q_{\rm R} \right)
               \left( \bar q_{\rm R} d_{\rm L} \right)
\, ,
\label{eq:2.3}
\end{eqnarray}
with similar expressions for $\widetilde{Q}_{5}$ and $\widetilde{Q}_{7}$.
Here $q_{R,L} = \frac{1}{2} (1\pm\gf) q$. The Fierz conjugates of the
$(V-A)\otimes (V-A)$ penguin operators $Q_{3}, Q_{4}, Q_{9}$ and
$Q_{10}$, to be denoted by $\widetilde{Q}_{i}, \; i=3,4,9,10$, can be
found in analogy to \eqn{eq:2.2}.

\subsection{Classification of Diagrams}

In order to calculate the anomalous dimension matrices $\gs$
and $\gss$, one has to insert the operators of eq.
\eqn{eq:2.1} in appropriate four--point functions and extract $1/\eps$
divergences.  The precise relation between the $1/\eps$ divergences in
one-- and two--loop diagrams and the one-- and two--loop anomalous
dimension matrices will be given in the following subsection. Here, let
us only recall that insertion of any of the operators of eq.
\eqn{eq:2.1} into the diagrams discussed below results in a linear
combination of the operators $Q_{i}$.  A row in the anomalous dimension
matrix corresponding to the inserted operator can then be obtained from
the coefficients in the linear combination in question.

There are three basic ways a given operator can be inserted into a
four--point function. They are shown in fig.~1, where the cross denotes
the interaction described by a current in a given operator of eq.
\eqn{eq:2.1}, and the wavy line denotes a gluon.  We shall refer to the
insertions of fig.~\ref{fig:1}(a) as ``current--current'' insertions.
The insertions of fig.~\ref{fig:1}(b) and (c) will then be called
``penguin insertions'' of type 1 and type 2 respectively.

The complete list of the diagrams necessary for one-- and two--loop
calculations is given in figs.~2--7. At the one--loop
level one has three current--current diagrams, fig.~2, to be
denoted as in ref. \cite{burasweisz:90} by $D_{1}$ -- $D_{3}$ and
one penguin diagram of each type, fig.~3, to be denoted by
$P_{0}^{(1)}$ and $P_{0}^{(2)}$ for type 1 and type 2 insertion
respectively. At the two--loop level there are 28 current--current diagrams
shown in fig.~4, to be denoted by $D_{4}$ -- $D_{31}$, and 14
penguin diagrams of each type shown in fig.~5, to be denoted by
$P_{1}^{(1)}$ -- $P_{14}^{(1)}$ and $P_{1}^{(2)}$ -- $P_{14}^{(2)}$
respectively.  The diagrams obtained from the ones given here by
left-right reflections have not been shown. In the case of
current--current diagrams also up-down reflections have to be
considered.  In fig.~6, we show two penguin diagrams which have no
$1/\eps$ divergence and hence do not contribute to the anomalous
dimensions, and in fig.~7, some examples of penguin diagrams which
vanish identically in dimensional regularization are given.

\subsection{Basic Formulae for the Anomalous Dimensions}

The anomalous dimensions of the operators $Q_{i}$, calculated in the
$\overline{MS}$ scheme, are obtained from the $1/\eps$ divergences of
the diagrams of figs.~2 -- 5 and from the $1/\eps$ divergence of
the quark wave function renormalization. Let us denote by $\vec{Q}$ a
column vector composed of the operators $Q_{i}$. Then
\begin{equation}
\hg(g) \; = \; \hat{Z}^{-1} \, \mu \,\frac{d}{d\mu} \,
\hat{Z},
\qquad {\rm with} \qquad
\vec{Q}^{B} \; = \; \hat{Z} \, \vec{Q} \,,
\label{eq:2.4}
\end{equation}
where $\vec{Q}^{B}$ stands for bare operators. Working in $D=4-2\,\eps$
dimensions, we can expand $\hat{Z}$ in inverse powers of $\eps$ as
follows
\begin{equation}
\hat{Z} \; = \; \hat{1} + \sum_{k=1}^{\infty} \frac{1}{\varepsilon^{k}}\,
\hat{Z}_{k}(g) \,.
\label{eq:2.5}
\end{equation}
where $g$ is the QCD coupling constant.
Inserting (\ref{eq:2.5}) into (\ref{eq:2.4}) one derives a useful formula
\begin{equation}
\hg(g) \; = \; -\,2\,g^{2}\,\frac{\partial\hat{Z}_{1}(g)}
{\partial g^{2}} \,.
\label{eq:2.6}
\end{equation}
Let us next denote by $\Gamma^{(4)}(\vec{Q})$ and
$\Gamma_{B}^{(4)}(\vec{Q}^{B})$ the renormalized and the bare four--quark
Green functions with operator $\vec{Q}$ insertions. Strictly speaking
$\Gamma^{(4)}(\vec{Q})$ and $\Gamma_{B}^{(4)}(\vec{Q}^{B})$ are matrices,
because the insertion of a single operator in a given diagram results in
a linear combination of operators.

At the one--loop level $\Gamma_{B}^{(4)}$ is obtained by evaluating the
diagrams of figs.~2 and 3. At the two--loop level it is found by evaluating
the diagrams of figs.~4 and 5 and subtracting the corresponding two--loop
counter terms. For the results of tabs. 1 -- 4, we have made these
subtractions diagram by diagram. Next
\begin{equation}
\Gamma^{(4)}(\vec{Q}) \; = \; \hat{Z}_{\psi} \hat{Z}^{-1} \,
\Gamma_{B}^{(4)}(\vec{Q}^{B}) \,,
\label{eq:2.7}
\end{equation}
where $\hat{Z}_{\psi}$ is a matrix which represents the renormalization
of the four quark fields. In the case of pure QCD it is a diagonal
matrix with all the diagonal elements being equal to
$Z^{{(\psi)}^{2}}$, where $Z^{(\psi)}$ is the quark wave function
renormalization, defined by
\begin{equation}
\psi^{B} \; = \; {Z^{(\psi)}}^{1/2} \, \psi \,.
\label{eq:2.8}
\end{equation}

We expand $\Gamma_{B}^{(4)}$ and $Z^{(\psi)}$ in inverse powers of
$\eps$ as follows
\begin{eqnarray}
Z^{(\psi)} & = & 1 + \sum_{k=1}^{\infty} \frac{1}{\varepsilon^{k}} \,
Z^{(\psi)}_{k}(g) \, , \label{eq:2.9} \\
\Gamma_{B}^{(4)}(Q_{i}) & = & \hat{1} +
\sum_{k=1}^{\infty} \frac{1}{\varepsilon^{k}} \, Z^{(\Gamma)}_{k}(g,Q_{i})
\; + \; {\rm finite} \, , \label{eq:2.9a}
\end{eqnarray}
where
\begin{equation}
Z^{(\psi)}_{1}(g) \; = \;
a_{1} \,\frac{g^{2}}{16\pi^{2}} +
a_{2} \,\frac{g^{4}}{(16\pi^{2})^{2}} +
{\cal O}(g^{6}) \, ,
\label{eq:2.10a}
\end{equation}
and
\begin{equation}
Z^{(\Gamma)}_{1}(g,Q_{i}) \; = \;
\sum_{j=1}^{10} \left[\,
\left(b_{1}\right)_{ij} \frac{g^{2}}{16\pi^{2}} +
\left(b_{2}\right)_{ij} \frac{g^{4}}{(16\pi^{2})^{2}} +
{\cal O}(g^{6})\, \right] \; .
\label{eq:2.11a}
\end{equation}

Demanding $\Gamma^{(4)}$ to be finite, we find $\hat{Z}_{1}$, and,
using (\ref{eq:2.6}), the basic formulae for the one-- and two--loop
anomalous dimension matrices,
\begin{eqnarray}
\big(\gs\big)_{ij} & = &
-\,2 \left[\,2\,a_{1}\,\delta_{ij} + \left(b_{1}\right)_{ij}\,\right]
\label{eq:2.10} \,, \\
\big(\gss\big)_{ij} & = &
-\,4 \left[\,2\,a_{2}\,\delta_{ij} + \left(b_{2}\right)_{ij}\,\right]
\label{eq:2.11} \,.
\end{eqnarray}

\subsection{Renormalization Scheme Dependence of $\gss$}

The two--loop anomalous dimension matrices depend on the renormalization
scheme and in particular on the treatment of $\gamma_{5}$ in $D\not=4$
dimensions. In ref. \cite{burasetal:92a}, we have derived a relation
between $(\gss)_{a}$ and $(\gss)_{b}$ calculated
in two different renormalization schemes $a$ and $b$,
\begin{equation}
\big(\gss\big)_{b} \; = \; \big(\gss\big)_{a} +
\left[ \Delta\rs, \gs \right] +
2 \,\beta_{0} \Delta\rs
\qquad {\rm with} \qquad
\Delta\rs \; = \; (\rs)_{b} - (\rs)_{a} \, ,
\label{eq:2.12}
\end{equation}
where $\beta_{0}$ is the leading coefficient in the expansion of the
$\beta$--function
\begin{equation}
\beta(g) \; = \; -\,\beta_{0}\,\frac{g^{3}}{16\pi^{2}} -
                    \beta_{1}\,\frac{g^{5}}{(16\pi^{2})^{2}} - \cdots \, ,
\label{eq:2.13}
\end{equation}
and the $(\rs)_{i}$ are defined by
\begin{equation}
\langle \vec{Q} \rangle_{i} \; = \;
\left[\, \hat{\bf 1} + \frac{g^{2}}{16\pi^{2}} \,(\rs)_{i} \,\right]
\langle \vec{Q}^{(0)} \rangle \; .
\label{eq:2.14}
\end{equation}
Here, $\vec{Q}^{(0)}$ is a tree--level matrix element and $\langle
\vec{Q} \rangle_{i}$ denotes renormalized one--loop matrix elements
calculated in the scheme $i$. The matrices $(\rs)_{i}$ are obtained
by calculating the finite terms in the one--loop diagrams of figs.~2 and 3.

Relation \eqn{eq:2.12} is very useful as it allows to test compatibility
of the two--loop anomalous dimension matrices calculated in two different
renormalization schemes. It also plays a central role in demonstrating
the scheme independence of physical quantities \cite{burasetal:92a}.

\subsection{Collection of Useful Results}

Let us recall the values for $\beta_{0}$, $\beta_{1}$, $a_{1}$, and
$a_{2}$,
\begin{equation}
\beta_{0} \; = \; \frac{11}{3} N - \frac{2}{3} f \, ,
\qquad \qquad
\beta_{1} \; = \; \frac{34}{3} N^{2} - \frac{10}{3} N f - 2\,C_{F} f \, ,
\label{eq:2.15}
\end{equation}
\begin{equation}
a_{1} \; = \; -\,C_{F} \, ,
\qquad \qquad
a_{2} \; = \;
C_{F} \left[\,\frac{3}{4}\,C_{F}-\frac{17}{4} N+\frac{1}{2} f\,\right] \, ,
\label{eq:2.16}
\end{equation}
where $C_{F} = (N^{2}-1)/2N$.
Here, $N$ is the number of colours and $f$ the number of active
flavours. All four quantities in eqs. \eqn{eq:2.15} and \eqn{eq:2.16}
are common to the NDR and HV scheme. Moreover $\beta_{0}$ and $\beta_{1}$
are gauge independent. The coefficients $a_{1}$ and $a_{2}$ given here
correspond to the Feynman gauge.

In the course of the discussion of the anomalous dimensions of the penguin
operators $Q_{6}$ and $Q_{8}$ it will be useful to have at hand the anomalous
dimension of the mass operator $\bar\psi \psi$,
\begin{equation}
\gamma_{m} \; = \; \gamma_{m}^{(0)} \,\frac{g^{2}}{16\pi^{2}} +
\gamma_{m}^{(1)} \,\frac{g^{4}}{(16\pi^{2})^{2}} + \cdots \, .
\label{eq:2.18}
\end{equation}
To two loops, the anomalous dimensions $\gamma_{m}^{(0)}$ and
$\gamma_{m}^{(1)}$ are given by
\begin{equation}
\gamma_{m}^{(0)} \; = \; -\,6\,C_{F}
\qquad {\rm and} \qquad
\gamma_{m}^{(1)} \; = \; -\,C_{F} \left[\,
3\ C_{F} + \frac{97}{3} N - \frac{10}{3} f \,\right] \, ,
\label{eq:2.19}
\end{equation}
for both schemes \cite{tarrach:81}.

Finally let us recall that for the anomalous dimension of the weak current,
\begin{equation}
\gamma_{J} \; = \; \gamma_{J}^{(0)} \frac{g^{2}}{16\pi^{2}} +
\gamma_{J}^{(1)} \frac{g^{4}}{(16\pi^{2})^{2}} + \cdots \, .
\label{eq:2.21}
\end{equation}
the one-loop coefficient $\gamma_{J}^{(0)}$ vanishes in both schemes.
However, at the two--loop level in the HV scheme, $\gamma_{J}^{(1)}$
is found to be non--vanishing
\cite{burasweisz:90},
\begin{equation}
\gamma_{J}^{(1)} \; = \; \left\{
\begin{array}{ll}
0                   &\qquad {\rm NDR} \\
4\,\beta_{0}\,C_{F} &\qquad {\rm HV}
\end{array} \right. \, .
\label{eq:2.22}
\end{equation}
In the HV scheme the axial current attains a finite renormalization if
one subtracts minimally, as we did, and one gets a mixing between
chiral components. This is not so aesthetic and on hindsight it would
have been more elegant to redefine the currents non-minimally to retain
standard normalization. We do not proceed along this line but in our
computation the mixing reappears in the non-vanishing of the two--loop
coefficient $\gamma_{J}^{(1)}$. However, since the one-loop matrix
elements $\hat{r}$ also change correspondingly, the final physical
quantities are normalized properly and scheme independent in the end.

\newsection{One--Loop Results}

\subsection{Current--Current Contributions to $\gs$}

The contributions of diagrams $D_{1}$ -- $D_{3}$ of fig.~2 to
the matrix $\gs$ have a very simple structure. The mixing
between different operators can be divided into five blocks, each
containing two operators,
\begin{equation}
\left( Q_{k}, Q_{k+1} \right) \, ,
\qquad \qquad
k = 1, 3, 5, 7, 9 \, ,
\label{eq:3.1}
\end{equation}
with no mixing between different blocks. Moreover, the mixing among
$(V-A)\otimes (V-A)$ operators $(k=1,3,9)$ is described by a universal
$2\times 2$ matrix.  Using eq. \eqn{eq:2.10}, one finds
\begin{equation}
\brackcc{ \gs(Q_{1}, Q_{2}) } \; = \;
\left( \begin{array}{cc}
-\,6/N &    6   \\
   6   & -\,6/N
\end{array} \right) \, ,
\label{eq:3.2}
\end{equation}
with identical results for $(Q_{3},Q_{4})$ and $(Q_{9},Q_{10})$.

Similarly,
\begin{equation}
\brackcc{ \gs(Q_{5}, Q_{6}) } \; = \;
\left( \begin{array}{cc}
 6/N & -\,6    \\
 0   & -\,6 \,(N^{2}-1)/N
\end{array} \right) \, ,
\label{eq:3.3}
\end{equation}
with identical result for $(Q_{7},Q_{8})$.

Comparing with eq. \eqn{eq:2.19}, we note that
\begin{equation}
\brackcc{ \gamma_{66}^{(0)} } \; = \;
\brackcc{ \gamma_{88}^{(0)} } \; = \; 2 \,\gamma_{m}^{(0)} \, ,
\label{eq:3.4}
\end{equation}
showing that the one--loop anomalous dimensions of the operators
$Q_{6}$ and $Q_{8}$ are twice the one--loop anomalous dimension of
the mass operator, provided only current--current diagrams
$D_{1}$ -- $D_{3}$ are taken into account. It should be pointed out that
this result is valid for an arbitrary number of colours $N$, but
remains only true in the Large-$N$ limit when penguin insertions
are taken into account \cite{bardeen:87,bardeen:87a}.

\subsection{Penguin Contributions to $\gs$}

The contributions of the diagrams $P_{0}^{(1)}$ and $P_{0}^{(2)}$ of
fig.~3 to the matrix $\gs$ have also a very simple
structure:
\begin{itemize}
\item
The insertions of $Q_{1}$, $Q_{5}$, and $Q_{7}$ into the diagrams
$P_{0}^{(1)}$ and $P_{0}^{(2)}$ vanish. This follows either from
colour conservation, or flavour conservation, or finally from the Dirac
structure, as one can easily verify by inspecting the diagrams of
fig.~3. Thus, the corresponding rows in $\brackp{\gs}$
vanish,
\begin{equation}
\brackp{ \gs(Q_{1}) } \; = \;
\brackp{ \gs(Q_{5}) } \; = \;
\brackp{ \gs(Q_{7}) } \; = \; 0 \, .
\label{eq:3.5}
\end{equation}
\item
The insertion of any of the remaining operators into the
diagrams of fig.~3 results always into a unique linear
combination of the QCD penguin operators $Q_{3}$ -- $Q_{6}$, multiplied
by an overall factor characteristic for the inserted operator.
Denoting by
\begin{equation}
P \; = \; \left( 0, \,0, \,-\,\frac{1}{N}, \,1, \,-\,\frac{1}{N}, \,1,
               \,0, \,0, \,0, \,0 \right)
\label{eq:3.6}
\end{equation}
the row vector in the space ($Q_{1}$ -- $Q_{10}$), the
non-vanishing elements of $\gs$, coming from  the diagrams
of fig.~3, are as follows,
\begin{equation}
\begin{array}{rclrcl}
\brackp{ \gs(Q_{2}) } & = & \frac{2}{3} \, P \, , &
\brackp{ \gs(Q_{3}) } & = &
\phantom{-} \,\frac{4}{3} \, P \, , \\
\brackp{ \gs(Q_{4}) } & = & \frac{2}{3} \, f P \, , &
\brackp{ \gs(Q_{6}) } & = &
\phantom{-} \,\frac{2}{3} \, f P \, , \\
\brackp{ \gs(Q_{8}) } & = &
\frac{2}{3} \big( u - \frac{d}{2} \big) \, P \, , &
\brackp{ \gs(Q_{9}) } & = & - \,\frac{2}{3} \, P \, , \\
\brackp{ \gs(Q_{10}) } & = &
\frac{2}{3} \big( u - \frac{d}{2} \big) \, P \, , & & &
\end{array}
\label{eq:3.7}
\end{equation}
where $u$ and $d$ denote the number of up-- and down--quark flavours
respectively ($u + d = f$). In obtaining \eqn{eq:3.7}, eq.
\eqn{eq:2.10} has been used with $a_{1}$ omitted, since this
contribution has already been included in \eqn{eq:3.2} and
\eqn{eq:3.3}.
\end{itemize}

Combining the results of eqs. \eqn{eq:3.2}, \eqn{eq:3.3}, \eqn{eq:3.5},
and \eqn{eq:3.7}, we find the known complete one--loop anomalous
dimension matrix $\gs$. For completeness this matrix has been given
explicitly in appendix A.

It should be stressed, that these results do not depend on the
renormalization scheme for $Q_{i}$ and are also valid for
$\widetilde{Q}_{i}$. This is no longer the case for the finite terms in the
diagrams of figs.~2 and 3, to which we now turn our attention.

\subsection{Current--Current Contributions to $\drs$}

The contribution to the matrix $\rs$ defined in eq. \eqn{eq:2.14},
resulting from the diagrams $D_{1}$ -- $D_{3}$ of fig.~2.
\begin{itemize}
\item
depends on the renormalization scheme for $Q_{i}$, but
\item
does not depend on whether $Q_{i}$, or their Fierz conjugates
$\widetilde{Q}_{i}$, are inserted in these diagrams.
\end{itemize}
The general structure of $\brackcc{\rs}$ is the same as in the case of
$\brackcc{\gs}$ presented in section~3.1.

Since $\hat{r}$ is generally gauge dependent and dependent on the
infrared structure of the theory we give here only results for
\begin{equation}
\drs \equiv \rshv - \rsndr
\label{eq:3.7a}
\end{equation}
which are free from these dependences. The results for $\rs$ in the
Landau--gauge have been given in ref.~\cite{burasetal:92a}.

We find
\begin{equation}
\brackcc{ \drs(Q_{1}, Q_{2}) } \; = \;
\left( \begin{array}{cc}
2 N - 4/N & 2 \\
2         & 2 N - 4/N
\end{array} \right) \, ,
\label{eq:3.8}
\end{equation}
with identical results for $(Q_{3}, Q_{4})$ and  $(Q_{9}, Q_{10})$.
Similarly,
\begin{equation}
\brackcc{ \drs(Q_{5}, Q_{6}) } \; = \;
\left( \begin{array}{cc}
2 N - 8/N & 6 \\
4         & 4 N - 8/N
\end{array} \right) \, ,
\label{eq:3.9}
\end{equation}
with identical result for $(Q_{7}, Q_{8})$.

\subsection{Penguin Contributions to $\drs$}

Turning our attention to the finite parts of the penguin diagrams of
fig.~3, it should be first noted that at the one loop level
one does not face the problem of the evaluation of ${\rm Tr}
[\gf\gamma_{\mu}\gamma_{\nu}\gamma_{\lambda}\gamma_{\omega}]$ in
$D \not= 4$ dimensions. Consequently, the diagram $P_{0}^{(1)}$ can be
calculated in the NDR renormalization scheme without any difficulties.
Nevertheless, different results are obtained for the NDR and HV scheme.

The reason that no problem with closed fermion loops arises at this
level is as follows. Although at the intermediate state ${\rm
Tr}(\gamma_\tau \gamma_5 \gamma_\nu \gamma_\lambda \gamma_\mu)$ appears
it is contracted with $g_{\mu\nu}$ and $q_\mu q_\nu$ where $q$ is the
momentum of the gluon. Thus actually only the trace ${\rm
Tr}(\gamma_\tau \gamma_5 \gamma_\lambda)$ has to be calculated, which
however is zero.

The structure of the results is similar to the ones given in
section~3.2.  In addition to the above mentioned scheme dependence, in
the case of the NDR, but not the HV scheme, the results for the
insertions of the operators $Q_{2}$, $Q_{3}$, $Q_{4}$, $Q_{9}$, and
$Q_{10}$ depend on whether these operators are taken in the basic form
of eq. \eqn{eq:2.1} or in the Fierz conjugate form
$\widetilde{Q}_{i}$.  In the case of pure QCD corrections the results
for $Q_{1}$, and for the $(V-A) \otimes (V + A)$ operators $Q_{5}$ --
$Q_{8}$, do not depend on the form used. These properties will play an
important role in section~5.

Beginning with the basis of eq. \eqn{eq:2.1}, we find that the
insertions of $Q_{1}$, $Q_{5}$, and $Q_{7}$ into the diagrams
$P_{0}^{(1)}$ and $P_{0}^{(2)}$ vanish in both renormalization schemes
considered. Thus the corresponding rows in $\brackp{\rsndr}$ and
$\brackp{\rshv}$ vanish,
\begin{equation}
\brackp{ \rsndr(Q_{1}) } \; = \;
\brackp{ \rsndr(Q_{5}) } \; = \;
\brackp{ \rsndr(Q_{7}) } \; = \; 0 \, ,
\label{eq:3.12}
\end{equation}
with an identical result for HV scheme.

The remaining insertions are non-zero but $\brackp{\drs}$ turns out to
vanish for $Q_4$, $Q_6$, $Q_8$ and $Q_{10}$:
\begin{equation}
\brackp{ \drs(Q_{4}) } \; = \;
\brackp{ \drs(Q_{6}) } \; = \;
\brackp{ \drs(Q_{8}) } \; = \;
\brackp{ \drs(Q_{10}) } \; = \; 0 \, .
\label{eq:3.12a}
\end{equation}
For the remaining operators we find then
\begin{equation}
\begin{array}{rcl}
\brackp{ \drs(Q_{2}) } & = & -\,\frac{1}{3} \, P \, , \\
\brackp{ \drs(Q_{3}) } & = & -\,\frac{2}{3} \, P \, , \\
\brackp{ \drs(Q_{9}) } & = & \phantom{-} \,\frac{1}{3} \, P \, ,
\end{array}
\label{eq:3.13}
\end{equation}
where the vector $P$ has been defined in \eqn{eq:3.6}.

Since, as stated above, $Q_2$ and $\tilde{Q}_2$ insertions give
different results in the NDR scheme, we find that the $\rsndr$ calculated
in the bases \eqn{eq:2.1} and \eqn{eq:2.2} differ. One obtains
\begin{equation}
\widetilde{\Delta}\rsndr \; \equiv \;
\rsndr[\ref{eq:2.2}] - \rsndr[\ref{eq:2.1}] \; = \;
\left( \begin{array}{c}
0 \\ -\frac{1}{3} P \\ 0 \\ \vdots \\ 0
\end{array} \right) \, .
\label{eq:3.16}
\end{equation}
This result will play an important role in section~5.

\newsection{Current--Current Contributions to the Two--Loop Anomalous
Dimension Matrix}

\subsection{$(V-A)\otimes (V-A)$ Operators}

The generalization of the one--loop matrix of eq. \eqn{eq:3.2} to
two--loop level can be done using the results of ref.
\cite{burasweisz:90}, where the eigenvalues of the $2 \times 2$
submatrix  $(Q_{1}, Q_{2})$ have been calculated in the NDR and HV
schemes. Since also at two--loop level the corresponding submatrix is
symmetric with equal entries on the diagonal, one has
\begin{eqnarray}
\brackcc{ \gamma_{11}^{(1)} } = &
\brackcc{ \gamma_{22}^{(1)} } = &
\frac{1}{2} \Big(\, \tg{+}^{(1)} + \tg{-}^{(1)} \,\Big) +
2 \,\gamma_{J}^{(1)} \, , \label{eq:4.1} \\
\brackcc{ \gamma_{12}^{(1)} } = &
\brackcc{ \gamma_{21}^{(1)} } = &
\frac{1}{2} \Big(\, \tg{+}^{(1)} - \tg{-}^{(1)} \,\Big)
\, , \label{eq:4.2}
\end{eqnarray}
with $\tg{\pm}^{(1)}$ given by formulae \eqn{eq:5.2} and
\eqn{eq:5.3} of ref. \cite{burasweisz:90} for NDR and HV schemes
respectively, and $\gamma_{J}^{(1)}$ in \eqn{eq:2.22} of the present
paper.

The details of the calculation of the diagrams $D_{4}$ -- $D_{31}$ of
fig.~4, with $(V-A) \otimes (V-A)$ insertions, can be found in
ref. \cite{burasweisz:90}, where a table of $1/\eps^{2}$ and $1/\eps$
singularities in the diagrams $D_{4}$ -- $D_{31}$ has been given. In
performing these calculations one has to take care of evanescent
operators which, although vanishing in $D = 4$, affect the two--loop
anomalous dimension matrix of physical operators $Q_{i}$ of eq.
\eqn{eq:2.1}. For this reason, the treatment of counter diagrams has to
be done with care. The discussion of evanescent operators can be found
in ref. \cite{burasweisz:90}, and a nice presentation is given also in
\cite{dugan:91}. Here it suffices to give only the projections on the space
of physical operators.  Denoting the evanescent operators generally by
$E_{\alpha\beta, \gamma\delta}$, with $\alpha$, $\beta$, $\gamma$,
$\delta$ being Dirac indices, the relevant projection for diagrams
involving $(V-A) \otimes (V-A)$ operators used in ref.
\cite{burasweisz:90} is given by
\begin{equation}
E_{\alpha\beta, \gamma\delta}  \;
\big(\gamma_{\tau} (1+\gamma_{5})\big)_{\beta\gamma} \,
\big(\gamma^{\tau} (1+\gamma_{5})\big)_{\delta\alpha} \; = \; 0 \, ,
\label{eq:4.3}
\end{equation}
where in the HV scheme $\gamma_{\tau}$ has to be taken in $D=4$.

Using eqs. \eqn{eq:4.1} and \eqn{eq:4.2}, together with the results of
ref. \cite{burasweisz:90}, we find in the case of the NDR scheme
\begin{equation}
\brackcc{\gssndr(Q_{1},Q_{2})} \; = \;
\left(\begin{array}{cc}
A_{\rm NDR} & B_{\rm NDR} \\
B_{\rm NDR} & A_{\rm NDR}
\end{array}\right) \, ,
\label{eq:4.4}
\end{equation}
where
\begin{equation}
A_{\rm NDR}  \; = \;  - \,\frac{22}{3} - \frac{2}{3} \frac{f}{N}
                      - \frac{57}{2} \frac{1}{N^2} \, ,
\qquad
B_{\rm NDR}  \; = \;  -\,\frac{19}{6} N + \frac{2}{3} f + \frac{39}{N} \, ,
\label{eq:4.6}
\end{equation}
with identical results for $(Q_{3}, Q_{4})$ and $(Q_{9}, Q_{10})$.

For the HV scheme we find
\begin{equation}
\brackcc{\gss(Q_{1},Q_{2}) } \; = \;
\left(\begin{array}{cc}
A_{\rm HV} & B_{\rm HV} \\
B_{\rm HV} & A_{\rm HV}
\end{array}\right) \, ,
\label{eq:4.7}
\end{equation}
where
\begin{equation}
A_{\rm HV} \; = \;
\frac{44}{3} N^{2} - \frac{8}{3} N f - \frac{110}{3} +
\frac{14}{3} \frac{f}{N} - \frac{57}{2} \frac{1}{N^2} \, ,
\qquad
B_{\rm HV} \; = \; \frac{23}{2} N - 2 f + \frac{39}{N}  \, ,
\label{eq:4.9}
\end{equation}
with identical results for $(Q_{3}, Q_{4})$ and $(Q_{9}, Q_{10})$.

\subsection{$(V-A)\otimes (V+A)$ Operators}

The calculation of the $(V-A) \otimes (V+A)$ insertions into the
diagrams $D_{4}$ -- $D_{31}$ proceeds in analogy with the one for
$(V-A)\otimes (V-A)$ operators. However, the projection on the space of
physical operators takes this time the following form,
\begin{equation}
E_{\alpha\beta, \gamma\delta} \;
(1-\gamma_{5})_{\beta\gamma} \,
(1+\gamma_{5})_{\delta\alpha} \; = \; 0 \, .
\label{eq:4.10}
\end{equation}

In order to check Fierz symmetry properties, it is instructive to
calculate simultaneously the insertions of the
$(1+\gf)\otimes(1-\gf)$ operators, e.g. $\widetilde{Q}_{6}$ of eq.
\eqn{eq:2.3}. In this case, the projection on the space of physical
operators takes the form
\begin{equation}
E_{\alpha\beta, \gamma\delta} \;
\big(\gamma_{\tau} (1+\gamma_{5})\big)_{\beta\gamma} \,
\big(\gamma^{\tau} (1-\gamma_{5})\big)_{\delta\alpha} \; = \; 0 .
\label{eq:4.11}
\end{equation}

The singular terms in the diagrams $D_{4}$ -- $D_{31}$ with $Q_{5}$ and
$\widetilde{Q}_{6}$ insertions in the NDR and HV scheme are collected in
table \ref{tab:1}.  The two--loop counter terms have been included
diagram by diagram. Also corrections to counter diagrams resulting from
the mixing with evanescent operators have been taken into account.
Colour factors have been omitted. They can be found in ref.
\cite{burasweisz:90}. The overall normalization is such that after the
inclusion of colour factors separate contributions to $(b_{2})_{ij}$ of
eq. \eqn{eq:2.11} are obtained. In the case of diagrams 29 -- 31, we have
included in table \ref{tab:1} the colour factors present in the gluon
self-energy
\begin{equation}
F_{1} \; = \; \frac{5}{3} N - \frac{2}{3} f \, ,
\qquad \qquad
F_{2} \; = \; \frac{31}{9} N - \frac{10}{9} f \, .
\label{eq:4.12}
\end{equation}
We make the following observations:
\begin{itemize}
\item
the $1/\eps^{2}$ singularities are the same in both schemes as expected,
\item
the $1/\eps$ singularities differ, which implies different two--loop
anomalous dimension matrices,
\item
the Fierz symmetry in diagrams $D_{4}$ -- $D_{31}$ is preserved in
both schemes. The Fierz conjugate pairs of diagrams are (4,6),
(7,9), (10,12), (13,15), (16,21), (17,20), (18,19), (22,24),
(25,27), and (29,31).
\end{itemize}

The remaining diagrams are self--conjugate. In ref.
\cite{burasweisz:90}, where the operators $\widetilde{Q}_{1}$ and $Q_{2}$
have been considered the Fierz conjugate diagrams with $\widetilde{Q}_{1}$
or $Q_{2}$ insertions had the same singular structure. Now, the Fierz
symmetry relates the $Q_{5}$ insertion of a given diagram to the
$\widetilde{Q}_{6}$ insertion of the Fierz--conjugate diagram. As seen in
table \ref{tab:1}, this relation is satisfied for all Fierz
conjugate diagrams: $Q_{5}$ and $\widetilde{Q}_{6}$ insertions give the
same result.  We should, however, emphasize that as in ref.
\cite{burasweisz:90}, also here this structure is only obtained after
the evanescent operators have been properly taken into account. Otherwise,
the Fierz symmetry in pairs (16,21), (17,20), (18,19) is broken and the
results for diagrams 5 and 23 are modified. Clearly, the fact that the
results in table \ref{tab:1} satisfy all these relations, is a good
check of our calculations.

Including colour factors, summing all diagrams $D_{4}$ -- $D_{31}$, and
using the formula \eqn{eq:2.11}, gives the current--current contribution
to the two--loop anomalous dimension matrix of $(V-A) \otimes (V+A)$
operators. For the NDR scheme we find
\begin{equation}
\brackcc{\gssndr(Q_{5},Q_{6})} \; = \;
\left(\begin{array}{cc}
C_{\rm NDR} & D_{\rm NDR} \\
E_{\rm NDR} & F_{\rm NDR}
\end{array}\right) \, ,
\label{eq:4.13}
\end{equation}
where
\begin{eqnarray}
C_{\rm NDR} & = &  \frac{137}{6} - \frac{22}{3} \frac{f}{N} +
\frac{15}{2} \frac{1}{N^{2}} \, ,
\qquad
D_{\rm NDR} \; = \; - \,\frac{100}{3} N + \frac{22}{3} f + \frac{3}{N} \, ,
\label{eq:4.15} \\
E_{\rm NDR} & = & - \,\frac{71}{2} N + 4 f - \frac{18}{N} \, ,
\qquad\phantom{--}
F_{\rm NDR} \; = \; 2 \,\gamma_{m}^{(1)} + \frac{89}{2} -
4 \,\frac{f}{N} + \frac{9}{N^{2}} \, ,
\label{eq:4.17}
\end{eqnarray}
with identical results for $(Q_{7}, Q_{8})$.
Here, $\gamma_{m}^{(1)}$ is the two--loop anomalous dimension of the
mass operator given in eq. \eqn{eq:2.19}.

For the HV scheme we find
\begin{equation}
\brackcc{\gsshv(Q_{5},Q_{6}) } \; = \;
\left(\begin{array}{cc}
C_{\rm HV} & D_{\rm HV} \\
E_{\rm HV} & F_{\rm HV}
\end{array}\right) \, ,
\label{eq:4.18}
\end{equation}
where
\begin{eqnarray}
C_{\rm HV} & = &
\frac{44}{3} N^{2} - \frac{8}{3} N f - \frac{71}{6} +
\frac{10}{3} \frac{f}{N} + \frac{15}{2} \frac{1}{N^{2}} \, ,
\quad
D_{\rm HV} \; = \; - \,\frac{40}{3} N - \frac{2}{3} f + \frac{3}{N} \, ,
\label{eq:4.20} \\
E_{\rm HV} & = & \frac{107}{6} N - \frac{4}{3} f - \frac{18}{N} \, ,
\quad
F_{\rm HV} \; = \; 2\,\gamma_{m}^{(1)} + \frac{88}{3} N^{2} -\frac{16}{3} N f -
\frac{229}{6} + \frac{20}{3} \frac{f}{N} + \frac{9}{N^{2}}\, , \nn
\end{eqnarray}
with identical results for $(Q_{7}, Q_{8})$.

\subsection{Properties and Test of Compatibility}

The matrices given in eqs. \eqn{eq:4.4}, \eqn{eq:4.7}, \eqn{eq:4.13},
and \eqn{eq:4.18} give the part of the two--loop anomalous dimension matrix
coming from the current--current diagrams of fig.~4. Let us
denote this part by $\brackcc{\gssndr}$ and
$\brackcc{\gsshv}$ for NDR and HV schemes
respectively. We note, that
\begin{itemize}
\item
these two matrices differ considerably from each other.
In particular all diagonal entries in the HV scheme increase as
$N^{2}$, whereas in the NDR scheme this is only the case for the
elements (6,6) and (8,8).
\item
Since $\gamma_{66}^{(0)}$ and $\gamma_{88}^{(0)}$ grow like $N$ for
large $N$ and $\gamma_{66}^{(1)}$ and $\gamma_{88}^{(1)}$ grow like
$N^2$ in both schemes considered, we find remembering $\as \sim 1/N$,
that these two elements approach non-vanishing constants in the
Large-$N$ limit. In the NDR scheme all other elements vanish in this
limit and this fact will not be changed by including penguin diagrams.
In the HV scheme all diagonal elements survive the Large-$N$ limit.
\item
the simple relation between $\gamma_{66}^{(0)}$, $\gamma_{88}^{(0)}$
and $\gamma_{m}^{(0)}$, given in \eqn{eq:3.4}, is violated at two--loop
level in both schemes.  However, in the NDR scheme it is recovered for
large $N$ \cite{pichderafael:91}.
\item
the vanishing elements (5,6) and (7,8) in eq. \eqn{eq:3.3}
become non--zero at the two--loop level.
\item
using the results of this and the previous section, it is interesting to
note that the relation \eqn{eq:2.12} is satisfied even if only
current-current contributions to $\gs$, $\rs$ and
$\gss$ are taken into account:
\end{itemize}
\begin{equation}
\brackcc{\gsshv } \; = \;
\brackcc{\gssndr} +
\left[ \Delta\brackcc{\rs}, \brackcc{\gs} \right] +
2 \,\beta_{0} \Delta\brackcc{\rs} \, ,
\label{eq:4.23}
\end{equation}
where
\begin{equation}
\Delta\brackcc{\rs} =
\brackcc{\rshv} - \brackcc{\rsndr}
\label{eq:4.24}
\end{equation}

This is easy to understand. Indeed, the two--loop
current--current diagrams of fig.~4 have subdiagrams of
current--current type only, and consequently the compatibility of
the two calculations should be verified within current--current
diagrams alone. The fact that this is indeed the case is a good
check of our results.

\newsection{Penguin Diagram Contributions to the Two-Loop Anomalous
Dimension Matrix}

\subsection{General Structure}

The calculation of the penguin diagram contributions to
$\gss$ can be considerably simplified by first analyzing the
general structure of these contributions. The two--loop penguin
diagrams are shown in fig.~5, where two types of insertions of
a given operator, b) and c) of fig.~1, have to be considered.
Yet, as we shall show now, the full matrix $\brackp{\gss}$ can
be obtained by calculating just the insertions of the operators
$Q_{1}$, $Q_{2}$, $\widetilde{Q}_{5}$, and $\widetilde{Q}_{6}$, which
moreover receive only contributions from the diagrams $P_{1}^{(2)}$ --
$P_{14}^{(2)}$, i.e. penguin diagrams without closed fermion loops.

We first note that the anomalous dimensions of the $(V-A) \otimes
(V-A)$ penguin operators $Q_{3}$, $Q_{4}$, $Q_{9}$, and $Q_{10}$ can be
expressed in terms of the anomalous dimensions of $Q_{1}$, $Q_{2}$,
$\widetilde{Q}_{1}$, and $\widetilde{Q}_{2}$ as follows,
\begin{eqnarray}
\brackp{\gss(Q_{3})} & = &
f \, \brackp{\gss(\widetilde{Q}_{1})} +
2 \, \brackp{\gss(Q_{2})} \, ,
\label{eq:5.1} \\
\brackp{\gss(Q_{4})} & = &
f \, \brackp{\gss(\widetilde{Q}_{2})} +
2 \, \brackp{\gss(Q_{1})} \, ,
\label{eq:5.2} \\
\brackp{\gss(Q_{9})} & = &
\Big(u - \frac{d}{2}\,\Big)\,\brackp{\gss(\widetilde{Q}_{1})} -
\brackp{\gss(Q_{2})} \, ,
\label{eq:5.3} \\
\brackp{\gss(Q_{10})} & = &
\Big(u - \frac{d}{2}\,\Big)\,\brackp{\gss(\widetilde{Q}_{2})} -
\brackp{\gss(Q_{1})} \, ,
\label{eq:5.4}
\end{eqnarray}
where the last terms in \eqn{eq:5.1} -- \eqn{eq:5.4} are only due to
internal down--quarks.

Similarly, the anomalous dimensions of the $(V-A)\otimes (V+A)$
operators $Q_{7}$ and $Q_{8}$ can be expressed in terms of those of
$Q_{5}$ and $Q_{6}$ respectively,
\begin{eqnarray}
\brackp{\gss(Q_{7})} & = &
\frac{(u-d/2)}{f} \, \brackp{\gss(Q_{5})} \, ,
\label{eq:5.5} \\
\brackp{\gss(Q_{8})} & = &
\frac{(u-d/2)}{f} \, \brackp{\gss(Q_{6})} \, .
\label{eq:5.6}
\end{eqnarray}

We next note that
$\widetilde{Q}_{1}$, $\widetilde{Q}_{2}$, $Q_{5}$, and $Q_{6}$ insertions
are non--zero only for $P_{1}^{(1)}$ -- $P_{14}^{(1)}$,
whereas $Q_{1}$, $Q_{2}$, $\widetilde{Q}_{5}$, and $\widetilde{Q}_{6}$
insertions are non--zero only for $P_{1}^{(2)}$ -- $P_{14}^{(2)}$.

In the HV scheme, there is no difficulty in evaluating the diagrams
$P_{1}^{(1)}$ -- $P_{14}^{(1)}$ which involve closed fermion loops.
Consequently, in this scheme, eqs. \eqn{eq:5.1} -- \eqn{eq:5.6} can
directly be used to find the matrix $\brackp{\gsshv}$.
However, in the NDR scheme it is much easier to deal with the diagrams
$P_{1}^{(2)}$ -- $P_{14}^{(2)}$. We will now show that relation
\eqn{eq:2.12} can be used to reduce the full calculation to the latter
diagrams.

Let us first observe that eq. \eqn{eq:2.12} can also be used to relate
the two--loop anomalous dimensions obtained in the bases \eqn{eq:2.1}
and \eqn{eq:2.2}. Since these two bases differ only in the two first
operators, and the current--current anomalous dimensions of
$(Q_{1}, Q_{2})$ and $(\widetilde{Q}_{1}, \widetilde{Q}_{2})$
are the same, we find
\begin{equation}
\brackp{\gss[\eqn{eq:2.2}]} \; = \;
\brackp{\gss[\eqn{eq:2.1}]} +
\left[ \widetilde{\Delta}\rs, \gs \right] +
2 \,\beta_{0} \widetilde{\Delta}\rs \, ,
\label{eq:5.7}
\end{equation}
where
\begin{equation}
\widetilde{\Delta}\rs \; = \;
\left\{ \begin{array}{ll}
 \widetilde{\Delta}\rsndr &\qquad {\rm NDR} \\
          0            &\qquad {\rm HV}
\end{array} \right. \, ,
\label{eq:5.8}
\end{equation}
with $\widetilde{\Delta}\rsndr$ given in eq. \eqn{eq:3.16}. Eqs.
\eqn{eq:5.7} and \eqn{eq:5.8} allow to find
$\brackp{\gss(\widetilde{Q}_{1})}$ and
$\brackp{\gss(\widetilde{Q}_{2})}$, once
$\brackp{\gss(Q_{1})}$ and
$\brackp{\gss(Q_{2})}$ are known. In the NDR scheme
we then have
\begin{eqnarray}
\brackp{\gssndr(\widetilde{Q}_{1})} & = &
\brackp{\gssndr(Q_{1})} + 2 \,P \, ,
\label{eq:5.9} \\
\brackp{\gssndr(\widetilde{Q}_{2})} & = &
\brackp{\gssndr(Q_{2})}
\label{eq:5.10} \\
 & + &
\frac{2}{9} \,\left( 0, \,0, \,2 - \frac{2}{N^{2}}, \,-\,11 N + \frac{11}{N},
\,11 + \frac{16}{N^{2}}, \,-\,2\,N - \frac{25}{N}, \,0, \,0, \,0, \,0
\right)\, \nn ,
\end{eqnarray}
whereas in the HV scheme, the insertions of $Q_{1}$ and $Q_{2}$ are
equal to the insertions of $\widetilde{Q}_{1}$ and $\widetilde{Q}_{2}$,
respectively.

Next, we recall the important result of section~3: the finite pieces in
one--loop diagrams (the matrix $\rs$) did not depend on whether $Q_{5}$
and $Q_{6}$ or $\widetilde{Q}_{5}$ and $\widetilde{Q}_{6}$ had been used.
Since $(Q_{5}, Q_{6})$ and $(\widetilde{Q}_{5}, \widetilde{Q}_{6})$ form
closed sets under the renormalization due to current--current diagrams,
and the insertion of any of these four operators in one--loop penguin
diagrams give always linear combinations of $Q_{3}$ -- $Q_{6}$ it is
evident that
\begin{eqnarray}
\brackp{\gss(Q_{5})} & = &
\brackp{\gss(\widetilde{Q}_{5})} \, ,
\label{eq:5.11} \\
\brackp{\gss(Q_{6})} & = &
\brackp{\gss(\widetilde{Q}_{6})} \, ,
\label{eq:5.12}
\end{eqnarray}
in both HV and NDR schemes.

To summarize, we have shown that the full matrix
$\brackp{\gss}$ can be obtained by calculating just the
insertions of the operators $Q_{1}$, $Q_{2}$, $\widetilde{Q}_{5}$, and
$\widetilde{Q}_{6}$ into the diagrams $P_{1}^{(2)}$ -- $P_{14}^{(2)}$.

\subsection{Details of the Two--Loop Calculation}

So far, we have restricted our discussion to ``on--shell'' results. It
is well known however, that to renormalize operators one has to include
mixing with all operators having the same quantum numbers. In the
literature \cite{klugberg:75,joglekar:76,deans:78,espriu:83}, it has
become standard to divide these into 3 classes I, II$^{a}$, and
II$^{b}$, according to
\begin{itemize}
\item class I: gauge invariant operators that do not vanish by virtue
 of the equations of motion,
\item class II$^{a}$: gauge invariant operators that vanish by virtue
 of the equations of motion, and
\item class II$^{b}$: gauge non--invariant operators.
\end{itemize}
In our case, the situation is even slightly more complicated, since one
has to consider also mixing with evanescent operators, which vanish
when restricted to 4 dimensions.

The renormalized operators $Q_{i}$ of eq. \eqn{eq:2.1}, corresponding
to the unrenormalized operators $Q_{i}^{B}$, which belong to class I,
are given by
\begin{equation}
Q_{i} \; = \; \sum_{j=1}^{6} Z_{ij}^{-1} \,Q_{j}^{B} + \sum_{k=1}^{K}
p_{ik} \,G_{k} + E_{i} + N_{i} \, .
\label{eq:5.2.1}
\end{equation}
Here, $E_{i}$ are the evanescent operators, $N_{i}$ are in class
II$^{b}$, and $G_{i}$ are gauge invariant operators of dimension 6
that just involve two quark fields. There are $K=3$ independent
operators of this sort (up to total derivatives), having $(\bar sd)$
quantum numbers which arise in our computation up to two--loops,
\footnote{The operator
$\left( \bar s\big[D^{2},D_{\mu}\big]\gamma^{\mu}(1-\gf) \,d \,\right)$
doesn't enter at two--loops.}
\begin{eqnarray}
G_{1} & = & \left( \bar s \big[ D_{\mu},\,[D^{\mu},D_{\nu}]\big]
            \gamma^{\nu}(1-\gf) \,d \,\right) \, , \nn \\
G_{2} & = & \left( \bar s \big\{D^{2},D_{\mu}\big\}
            \gamma^{\mu}(1-\gf) \,d \,\right) \, , \label{eq:5.2.2} \\
G_{3} & = & \left( \bar s D_{\mu} D_{\nu} D_{\lambda} S^{\mu\nu\lambda}
            (1-\gf) \,d \,\right)  \, , \nn
\end{eqnarray}
where $D_{\mu}$ is the covariant derivative and
$S^{\mu\nu\lambda}=\gamma^{\mu}\gamma^{\nu}\gamma^{\lambda}-
\gamma^{\lambda}\gamma^{\nu}\gamma^{\mu}$.

The operator $G_{1}$ is proportional to the penguin operator $Q_{p}$
which occurs at one--loop,
\begin{equation}
Q_{p} \; = \; \frac{4}{g} \,\big( \bar s \gamma_{\mu}(1-\gf) \,d
              \,D_{\nu} F^{\mu\nu} \big) \, ,
\label{eq:5.2.3}
\end{equation}
and
\begin{equation}
G_{2} \, , \quad G_{1} + G_{3} \, , \quad {\rm and} \quad
G_{1} - Q_{4} - Q_{6} + \frac{1}{N}\,\big(\, Q_{3} + Q_{5} \,\big) \, ,
\label{eq:5.2.4}
\end{equation}
all belong to class II$^{a}$.

In arbitrary gauges it is known that class I operators generally mix
with class II operators, in particular with those of class II$^{b}$.
One can simplify the work considerably by using the background field
gauge \cite{abbott:81}, for which no such mixing with gauge non--invariant
operators occurs. In our calculation, we therefore used such a gauge
\footnote{We did not compute in an arbitrary covariant background field
gauge, but restricted ourselves to the Feynman background field gauge.},
and thus have $N_{i}=0$ in eq. \eqn{eq:5.2.1}. However this choice only
affects the diagrams with triple gluon vertices, $P_{6}$ and $P_{7}$
of fig.~5.

In order to include mixing with the operators of class II$^{a}$,
\eqn{eq:5.2.2}, we have to calculate the operator insertions into
four--quark Green functions with arbitrary external momenta. This
results in additional momentum dependent operators which reflect
the presence of class II$^{a}$ operators. The various tensor products,
leaving out the external quark fields, which appear in the calculation
are as follows,
\begin{eqnarray}
V_{1} & = & \gamma_{\mu}(1-\gf) \; \otimes \; \gamma^{\mu} \, , \nn \\
V_{2} & = & \Slash q\,(1-\gf) \; \otimes \; \Slash q \,
\frac{1}{q^{2}} \, , \nn \\
V_{3} & = & V_{1} \, \frac{p_{1}\,p_{2}}{q^{2}} \, , \label{5.2.5} \\
V_{4} & = & \big(\, \Slash p_{1}(1-\gf) \; \otimes \; \Slash p_{2}\, +
\Slash p_{2}(1-\gf) \; \otimes \; \Slash p_{1} \,\big) \,
\frac{1}{q^{2}} \, , \nn \\
V_{5} & = & S_{\mu\lambda\nu} \,p_{1}^{\mu} \,p_{2}^{\nu}(1-\gf) \;
\otimes \; \gamma^{\lambda} \, \frac{1}{q^{2}} \, , \nn
\end{eqnarray}
where $p_{1}$, $p_{2}$ are incoming (outgoing) quark momenta and
$q=p_{1}-p_{2}$. The operators $Q_{i}$ of eq. \eqn{eq:2.1} all
correspond to the structure $V_{1}$.

The procedure to calculate the diagrams of fig.~5 followed that described
in ref. \cite{burasweisz:90}. That is, we first classified all independent
integrals occuring in the diagrams and their singular parts, computed
for arbitrary external momenta. Then the contractions with the
corresponding $\gamma$--matrix structures, using the NDR rules and the
HV rules, were performed separately. For the NDR scheme this was done
by ``hand'' and by using the algebraic computer program TRACER
\cite{jamin:91}. For the HV scheme we used TRACER nearly exclusively
for the penguin diagrams.

The operator structures $V_{2}$ and $V_{4}$ already vanish going
on-shell, i.e. using the equations of motion, while the on--shell
projection of $V_{3}$ and $V_{5}$ is given by
\begin{equation}
V_{3} \; \longrightarrow \; -\,\frac{1}{2} \,V_{1}
\qquad {\rm and} \qquad
V_{5} \; \longrightarrow \; V_{1} \, .
\label{eq:5.2.6}
\end{equation}
As a check, the on-shell result was also calculated in the ordinary
Feynman gauge. Since only diagrams $P_{6}$ and $P_{7}$ are affected,
after using the equations of motion their sum should agree with the
result in the background field gauge. Up to evanescent contributions
this is indeed the case.

A second technicality of the calculation that we like to discuss in
somewhat more detail is the treatment of $\gf$.

There are basically two diagrams, namely diagrams~3 and 4 of
fig.~\ref{fig:5}, where one encounters problems with the traces
involving $\gf$.  There we meet the ambigous trace expression
\begin{equation}
U_{\sigma} \; = \; S^{\,\mu\nu\lambda}\,Tr\,\big[S_{\mu\nu\lambda}
\gamma_{\sigma}\gf\big]\,,
\label{eq:Usig}
\end{equation}
where $S_{\mu\nu\lambda}$ is defined as before. All other ambigous
expressions can be reduced to this one together with other
non--ambigous terms. Besides the treatment of this problem in NDR as
described in the previous section, we can attempt an Ansatz for
$U_{\sigma}$ to replace it by
\begin{equation}
U_{\sigma}\quad\longrightarrow\quad -\,96\,\Big(1+v\,\eps+\ldots
\Big)\,\gamma_{\sigma}\gf\,,
\label{eq:repl}
\end{equation}
with $v$ representing the ambiguity which is of ${\cal O}(\eps)$.

Although $v$ does appear at intermediate steps of the calculation of
diagrams 3 and 4, after inclusion of evanescent contributions, it cancells
out in the final expression. The result thus obtained agrees with the
result obtained through the procedure of sect.~5.1. This gives us
confidence that our treatment of NDR is correct.

We would like to point out that the replacement of eq. (\ref{eq:repl})
seems to be related to the procedure recently proposed by K\"orner et.al.
\cite{koerner:90}. We emphasize that in both works the scheme presented
is a prescription, which although shown to work upto two--loops for
certain processes, still needs to be proven consistent to all orders.

\subsection{Results}

The singular terms in the diagrams $P_{1}^{(2)}$ -- $P_{14}^{(2)}$ with
$Q_{2}$ and $\widetilde{Q}_{6}$ insertions in the NDR and HV scheme, are
collected in tables \ref{tab:2}--\ref{tab:4}. The two--loop counter
terms have been included diagram by diagram.  Colour factors have been
omitted. The overall normalization is such that after the inclusion of
colour factors separate contributions to $(b_{2})_{ij}$ of eq.~\eqn{eq:2.11}
are obtained. The singularites in tables~\ref{tab:2}--\ref{tab:4}
include also contributions from penguin diagrams obtained from
fig.~\ref{fig:5} by left-right reflections. It should be stressed that
these additional diagrams give generally different singularities than
the diagrams given explicitly in fig.~\ref{fig:5}. As usual, the result
of a given penguin diagram contains two Dirac structures,
\begin{equation}
V_{LR} \; = \; \gamma_{\mu} \left( 1 - \gamma_{5} \right) \otimes
               \gamma^{\mu} \left( 1 + \gamma_{5} \right) \, ,
\qquad \enspace
V_{LL} \; = \; \gamma_{\mu} \left( 1 - \gamma_{5} \right) \otimes
               \gamma^{\mu} \left( 1 - \gamma_{5} \right) \, .
\label{eq:5.13}
\end{equation}

In diagrams $P_{1}^{(2)}$ -- $P_{10}^{(2)}$ the singular factors
multiplying $V_{LR}$ and $V_{LL}$ are the same, but they differ in the
case of $P_{11}^{(2)}$ -- $P_{14}^{(2)}$. We observe, that diagrams
$P_{1}^{(2)}$, $P_{9}^{(2)}$, and $P_{10}^{(2)}$ do not contribute to
the anomalous dimensions.

The singular terms for $Q_{2}$ and $\widetilde{Q}_{6}$ given in tables
\ref{tab:2}--\ref{tab:4} are also valid for $Q_{1}$ and $\widetilde{Q}_{5}$
respectively. However, since $Q_{1}$ and $\widetilde{Q}_{5}$ are in the
colour non--singlet forms, whereas $Q_{2}$ and $\widetilde{Q}_{6}$ are in
the singlet form, the colour factors differ. In fact, it turns out that in
the case of the insertion of $Q_{1}$ and $\widetilde{Q}_{5}$ only the colour
factors in diagrams 3, 4, 11, and 12 are non--vanishing, and only these
diagrams contribute to $\gss(Q_{1})$ and
$\gss(\widetilde{Q}_{5})$.

Including colour factors, using the results of tables
\ref{tab:2}--\ref{tab:4} and eq. \eqn{eq:2.11} without $a_{2}$
(taken already into account in the current--current contributions), we
obtain the expressions for $\brackp{\gss(Q_{1})}$,
$\brackp{\gss(Q_{2})}$,
$\brackp{\gss(\widetilde{Q}_{5})}$, and
$\brackp{\gss(\widetilde{Q}_{6})}$, from which the full
matrix $\brackp{\gss}$ can be found by means of formulae of
sect. 5.1.

In the NDR scheme we have
\begin{eqnarray}
\brackp{\gssndr(Q_{1})} & = & \left(
3 N - \frac{2}{3 N},\, - \,\frac{7}{3},\, - \,3 N + \frac{16}{3 N},\,
- \,\frac{7}{3} \right) \, ,
\label{eq:5.14} \\ \mvs
\brackp{\gssndr(Q_{2})} & = & \left(
-\,\frac{32}{27} + \frac{86}{27 N^{2}},\,\frac{176}{27} N - \frac{230}{27 N},\,
-\,\frac{122}{27} - \frac{94}{27 N^{2}},\, \frac{86}{27} N + \frac{130}{27 N}
\right) \, ,
\label{eq:5.15} \\ \mvs
\brackp{\gssndr(\widetilde{Q}_{5})} & = & \left(
-\,3 N + \frac{20}{3 N},\, -\,\frac{11}{3},\, 3 N + \frac{2}{3 N},\,
-\,\frac{11}{3} \right) \, f \, ,
\label{eq:5.16} \\ \mvs
\brackp{\gssndr(\widetilde{Q}_{6})} & = & \left(
-\,\frac{56}{27} - \frac{178}{27 N^{2}},\, -\,\frac{16}{27} N +
\frac{250}{27 N},\, \frac{70}{27} + \frac{74}{27 N^{2}},\,
\frac{110}{27} N - \frac{254}{27 N} \right)\,  f \, , \;\;\qquad
\label{eq:5.17}
\end{eqnarray}
where the vectors are in the $(Q_{3}, Q_{4}, Q_{5}, Q_{6})$ space.

In the HV scheme we have
\begin{eqnarray}
\brackp{\gsshv(Q_{1})} & = & \left(
3 N - \frac{4}{N},\, 1,\, - \,3 N + \frac{2}{N},\, 1 \right) \, ,
\label{eq:5.18} \\ \mvs
\brackp{\gsshv(Q_{2})} & = & \left(
-\,\frac{56}{27} + \frac{86}{27 N^{2}},\, \frac{110}{27} N -\frac{140}{27 N},\,
-\,\frac{128}{27} - \frac{58}{27 N^{2}},\, \frac{38}{27} N + \frac{148}{27 N}
\right) \, ,
\label{eq:5.19} \\ \mvs
\brackp{\gsshv(\widetilde{Q}_{5})} & = & \left(
- \,3 N + \frac{8}{3 N},\, \frac{1}{3},\, 3 N - \frac{10}{3 N},\, \frac{1}{3}
\right) \, f \, ,
\label{eq:5.20} \\ \mvs
\brackp{\gsshv(\widetilde{Q}_{6})} & = & \left(
-\,\frac{128}{27} - \frac{94}{27 N^{2}},\, \frac{20}{27} N +\frac{202}{27 N},\,
-\,\frac{38}{27} + \frac{86}{27 N^{2}},\, \frac{110}{27} N - \frac{158}{27 N}
\right)\,  f \, . \;\;\qquad
\label{eq:5.21}
\end{eqnarray}

Using next eqs. \eqn{eq:5.1} -- \eqn{eq:5.6} in conjunction with
\eqn{eq:5.9} -- \eqn{eq:5.12}, we find the full matrices
$\brackp{\gssndr}$ and $\brackp{\gsshv}$.

We observe that for large $N$ the contributions from penguin diagrams
to $\gss$ grow at most as $N$. Since $\as^2 \sim \ord(1/N^2)$ the full
contribution of penguin diagrams to the third term in eq.  \eqn{eq:1.1}
vanishes in the Large-$N$ limit. Yet for $N=3$ the penguin diagrams
play a considerable role in the numerical values for $\gss$.

\newsection{Full Two--Loop Anomalous Dimension Matrix $\gss$}

\subsection{Basic Result of this Paper}

Adding the current-current and penguin contributions to
$\gss$ found in sections~4 and 5 respectively, we obtain the
complete two--loop matrices $\gssndr$ and
$\gsshv$.

We first note that 48 elements of these matrices vanish in both
schemes. These are given by
\begin{equation}
\brackets{\gssndr}{ij} \; = \;
\brackets{\gsshv }{ij}  \; = \; 0
\label{eq:6.1}
\end{equation}
where
\begin{equation}
\begin{array}{lll}
\{ i = 3, \ldots, 10, \; j = 1,2 \} \, ;
& \qquad \enspace &
\{ i = 1, \ldots, 6, \; j = 7, \ldots, 10 \} \, ;
\\
\{ i = 7,8,  \; j = 9,10 \} \, ;
& \qquad \enspace &
\{ i = 9,10, \; j = 7,8 \}
\end{array}
\label{eq:6.2}
\end{equation}

The remaining 52 elements of each matrix are non--vanishing.  These
entries in the anomalous dimension matrix are given in
table~\ref{tab:5} for arbitrary number of colours and flavours.

For phenomenological applications we need only the results with $N=3$.
We give them in appendix C for an arbitrary number of flavours. It is
evident from this table that the two--loop $\ord(\as^2)$ corrections to
the anomalous dimensions of the operators $Q_i$ are substantial although
they are quite different for the two schemes considered.

\subsection{Compatibility of NDR and HV Results}
Using the results of table~5 and the one-loop results for $\gs$,
$\drs$ of section 3 we can test whether the compatibility
relation \eqn{eq:2.12} is satisfied.  This turns out to be indeed the
case which constitutes a test of our calculation.

\subsection{An Additional Test of the Calculation}

On general grounds, renormalizability relates the singularities
of degree $k$ in the $n$-th order of perturbation theory with
the singularities of degree $(k+1)$ in the $(n+1)$-th order. This
fact allows for an additional check of our computation by
considering the corresponding relation between the coefficients
of the $1/\eps^{2}$ singularities at two--loops and the
$1/\eps$ singularities at one--loop. Demanding the finiteness
of the renormalized Green function $\Gamma^{(4)}$, and the
anomalous dimension matrix $\hg$, a procedure analogous
to the derivation of eqs.~\eqn{eq:2.10} and \eqn{eq:2.11} leads to
\begin{equation}
\hG_{22} \; = \; -\,\frac{1}{8}\,\gs{}^{2} +
\Big(\,\frac{1}{4}\,\beta_{0}-a_{1}\Big)\gs+\Big[\,
\big(\beta_{0}+\delta_{0}\big)\,a_{1}-2a_{1}^{2}\,\Big]\,\hat{1}\,,
\label{eq:test}
\end{equation}
where $\hG_{22}$ is the coefficient of the $1/\eps^{2}$
singularity at ${\cal O}(\as^{2})$ of the bare Greens function
$\Gamma_{B}^{(4)}$, and $\delta_{0}$ is the leading order coefficient
of the renormalization group function associated with the gauge
parameter. In the Feynman gauge it is given by
\begin{equation}
\delta_{0} \; = \; -\,\frac{5}{3}N+\frac{2}{3}f\,.
\label{eq:del0}
\end{equation}
The values for $\beta_{0}$ and $a_{1}$ were already given in
eqs.~\eqn{eq:2.15} and \eqn{eq:2.16} respectively.

However, to perform the test we have to add the contributions of the
diagrams of fig.\ref{fig:6} to the result which can be obtained by means of
tables~\ref{tab:1}--\ref{tab:4}. Despite the fact that they have no $1/\eps$
divergence and hence do not contribute to the anomalous dimension
matrix, they contribute to $\hG_{22}$. Their contribution to
$\hG_{22}$ is given by
\begin{equation}
\Delta\hG_{22} \; = \; \frac{1}{2}\,\big(\beta_{0}+C_{F}\big)
\,\gs\,,
\label{eq:delga}
\end{equation}
where the term proportional to $\beta_{0}$ stems from the
renormalization of the gluon--propagator and the term proportional to
$C_{F}$ from the vertex renormalization. Let us recall that we work in
the background field gauge, where the $\beta$--function can be obtained
directly from the renormalization of the gluon propagator
\cite{abbott:81}.

Putting everything together, we find that the relation of
eq.~(\ref{eq:test}) is indeed satisfied. In addition, similar to the
test of the compatibility of different schemes, this relation could
also be split in a part solely originating from current--current
diagrams which is satisfied separately, and a relation involving mixing
between current--current and penguin diagrams. We will however not
elaborate on this any further.

\newsection{Summary}
We have presented the details and the explicit results of the
calculation of the two--loop anomalous dimension matrix $\ord(\as^2)$
involving current-current, QCD-penguin and electroweak-penguin operators.
Performing the calculation in two schemes for $\gf$, NDR and HV, we
have verified the compatibility of the anomalous dimension matrices
obtained in these two schemes. We have shown how the use of two
different operator bases allows to avoid a direct calculation of
penguin diagrams with closed fermion loops. This enabled us to find in
an unambiguous way the two--loop anomalous dimensions in the simplest
dimensional regularization scheme, the one with anticommuting $\gf$.
The results of this paper generalize our earlier paper
\cite{burasetal:92a} where electroweak penguin operators have not been
taken into account. The two--loop anomalous dimensions matrix presented
for the first time in ref.~\cite{burasetal:92a} and generalized in
section~6 of the present paper will play a central role in any analysis
of non-leptonic decays of hadrons which goes beyond the leading
logarithmic approximation. A consistent next-to-leading order analysis
involving electroweak penguin operators requires the calculation of
$\gse$, i.e.~ the two--loop anomalous dimension matrix $\ord(\aem\as)$.
A detailed account of this calculation is given in
\cite{burasetal:92c}. The full renormalization group analysis with the
anomalous dimension matrix of eq.~\eqn{eq:1.1} will be presented soon
in ref.~\cite{burasetal:92d}. There the numerical values for the Wilson
coefficient functions $C_i$ of the operators $Q_i$ including
next-to-leading order corrections can be found.


\vskip 1cm
\begin{center}
{\Large\bf Acknowledgement}
\end{center}
\noindent
We would like to thank many colleagues for a continuous encouragement
during this calculations. One of us (A.J.B.) would like to thank Bill
Bardeen and David Broadhurst for very interesting discussions.
M.E.L.~is grateful to Stefan Herrlich for a copy of his program {\tt
feynd} for drawing the Feynman diagrams in the figures and to Gerhard
Buchalla for stimulating discussions.


\newpage
\appendix{\LARGE\bf\noindent Appendices}

\newsection{One--Loop Anomalous Dimension Matrix $\gs$}

\bigskip


\hskip -1cm
$ \gs =
\left(
\begin{array}{cccccccccc}
{{-6}\over N} & 6 & 0 & 0 & 0 & 0 & 0 & 0 & 0 & 0 \\ \svs
6 & {{-6}\over N} & {{-2}\over {3 N}} & {2\over 3} & {{-2}\over {3 N}} &\
  {2\over 3} & 0 & 0 & 0 & 0 \\ \svs
0 & 0 & {{-22}\over {3 N}} & {{22}\over 3} & {{-4}\over {3 N}} & {4\over 3} &\
  0 & 0 & 0 & 0 \\ \svs
0 & 0 & 6 - {{2 f}\over {3 N}} & {{-6}\over N} + {{2 f}\over 3} & {{-2\
  f}\over {3 N}} & {{2 f}\over 3} & 0 & 0 & 0 & 0 \\ \svs
0 & 0 & 0 & 0 & {6\over N} & -6 & 0 & 0 & 0 & 0 \\ \svs
0 & 0 & {{-2 f}\over {3 N}} & {{2 f}\over 3} & {{-2 f}\over {3 N}} & {{-6\
  \left( -1 + {N^2} \right) }\over N} + {{2 f}\over 3} & 0 & 0 & 0 & 0 \\ \svs
0 & 0 & 0 & 0 & 0 & 0 & {6\over N} & -6 & 0 & 0 \\ \svs
0 & 0 & {{-2 \left( u-d/2 \right) }\over {3 N}} & {{2 \left(\
  u-d/2 \right) }\over 3} & {{-2 \left( u-d/2 \right)\
  }\over {3 N}} & {{2 \left( u-d/2 \right) }\over 3} & 0 & {{-6\
  \left( -1 + {N^2} \right) }\over N} & 0 & 0 \\ \svs
0 & 0 & {2\over {3 N}} & -{2\over 3} & {2\over {3 N}} & -{2\over 3} & 0 & 0 &\
  {{-6}\over N} & 6 \\ \svs
0 & 0 & {{-2 \left( u-d/2 \right) }\over {3 N}} & {{2 \left(\
  u-d/2 \right) }\over 3} & {{-2 \left( u-d/2 \right)\
  }\over {3 N}} & {{2 \left( u-d/2 \right) }\over 3} & 0 & 0 & 6\
  & {{-6}\over N}
\end{array}
\right) $


\bigskip


\newsection{Tables of Singularities and Two--Loop Anomalous Dimension
Matrix $\gss$}

Tables \ref{tab:1}--\ref{tab:4} list the $1/\eps^2$-- and
$1/\eps$--singularities of the various Feynman diagrams obtained in the
NDR and HV scheme. The values given already include diagram by diagram
two--loop counterterms. Colour factors are omitted and a common overall
factor $\as^2/(4\pi)^2$ is to be understood.

In table \ref{tab:1} the column labeled $D$ refers to the numbering of
current--current diagrams in figs.~\ref{fig:2} and \ref{fig:4}.
Similarly, the $P$-column in tables \ref{tab:2}--\ref{tab:4} refers to
penguin diagrams of figs.~\ref{fig:5}. In addition, in table
\ref{tab:1} the column labeled $M$ gives the multiplicity of each
diagram, which already has been included in the singularities quoted.
The singularites in tables~\ref{tab:2}--\ref{tab:4} include also
contributions from diagrams obtained from fig.~\ref{fig:5} by left-right
reflections.

\begin{table}[ht]
\caption[]{
\label{tab:1}}
\begin{center}
\begin{tabular}{|c|c||c|c|c||c|c|c|}
\hline
\multicolumn{2}{|c||}{} &
\multicolumn{3}{c||}{$Q_{5}$} & \multicolumn{3}{c|}{$\widetilde{Q}_{6}$} \\
\hline
$D$ & $M$ &
$ 1/\eps^{2} $ &
$\left( 1/\eps \right)_{\rm NDR}$ &
$\left( 1/\eps \right)_{\rm HV}$  &
$ 1/\eps^{2} $ &
$\left( 1/\eps \right)_{\rm NDR}$ &
$\left( 1/\eps \right)_{\rm HV}$  \\
\hline
\hline
 4 & 2  & -- 1  &   5/2 &   5/2 & -- 16 &    16 &    16 \\
 5 & 2  & -- 1  &   5/2 &   5/2 & -- 1  &   5/2 &   5/2 \\
 6 & 2  & -- 16 &    16 &    16 & -- 1  &   5/2 &   5/2 \\
 7 & 2  &    -- & -- 4  & -- 4  &    -- & -- 4  & -- 4  \\
 8 & 2  &    -- & -- 4  & -- 4  &    -- & -- 4  & -- 4  \\
 9 & 2  &    -- & -- 4  & -- 4  &    -- & -- 4  & -- 4  \\
10 & 4  & -- 2  &    3  & -- 1  & -- 8  & -- 8  & -- 16 \\
11 & 4  &    2  &    3  & -- 1  &    2  &    3  & -- 1  \\
12 & 4  & -- 8  & -- 8  & -- 16 & -- 2  &    3  & -- 1  \\
13 & 4  &    2  & -- 3  &    1  &    8  & -- 4  &    4  \\
14 & 4  & -- 2  & -- 3  &    1  & -- 2  & -- 3  &    1  \\
15 & 4  &    8  & -- 4  &    4  &    2  & -- 3  &    1  \\
16 & 4  &    2  & -- 9  & -- 1  &    8  & -- 16 &    8  \\
17 & 4  &    2  &    3  & -- 5  &    8  &    16 & -- 8  \\
18 & 4  & -- 8  &    8  &    -- & -- 8  & -- 8  &    -- \\
19 & 4  & -- 8  & -- 8  &    -- & -- 8  &    8  &    -- \\
20 & 4  &    8  &    16 & -- 8  &    2  &    3  & -- 5  \\
21 & 4  &    8  & -- 16 &    8  &    2  & -- 9  & -- 1  \\
22 & 1  & -- 1  &    -- &    -- & -- 16 &    -- &    -- \\
23 & 1  & -- 1  &    -- &    -- & -- 1  &    -- &    -- \\
24 & 1  & -- 16 &    -- &    -- & -- 1  &    -- &    -- \\
25 & 4  &    6  & -- 11 &    1  &    24 & -- 20 &    4  \\
26 & 4  & -- 6  & -- 7  &    5  & -- 6  & -- 7  &    5  \\
27 & 4  &    24 & -- 20 &    4  &    6  & -- 11 &    1  \\
28 & 4  &    -- &    -- &    -- &    -- &    -- &    -- \\
29 & 2  & --  & $\frac{3}{2} F_{1}$  & $-\,\frac{1}{2} F_{1}$  &
$-\,3 F_{1}$ & $3 F_{2} - F_{1}$  & $3 F_{2} - 5 F_{1}$  \\
30 & 2  & --  &  $\frac{3}{2} F_{1}$  & $-\,\frac{1}{2} F_{1}$  &
-- & $\frac{3}{2} F_{1}$  & $-\,\frac{1}{2} F_{1}$  \\
31 & 2  & $-\,3 F_{1}$  & $3 F_{2} - F_{1}$  & $3 F_{2} - 5 F_{1}$  &
-- & $\frac{3}{2} F_{1}$  & $-\,\frac{1}{2} F_{1}$  \\
\hline
\end{tabular}
\end{center}
\end{table}

\newpage
\begin{table}[h]
\caption[]{
\label{tab:2}}
\begin{center}
\begin{tabular}{|c||c|c|c||c|c|c|}
\hline
&
\multicolumn{3}{c||}{$Q_{2}$} & \multicolumn{3}{c|}{$\widetilde{Q}_{6}$} \\
\hline
$P$ &
$ 1/\eps^{2} $ &
$\left( 1/\eps \right)_{\rm NDR}$ &
$\left( 1/\eps \right)_{\rm HV}$  &
$ 1/\eps^{2} $ &
$\left( 1/\eps \right)_{\rm NDR}$ &
$\left( 1/\eps \right)_{\rm HV}$  \\
\hline
\hline
 1 &    2/3  &    --    &    --    &    2/3  &    --    &    --     \\
 2 &    2/3  & -- 19/9  & -- 13/9  &    2/3  & -- 13/9  & -- 13/9   \\
 3 &    --   & -- 17/9  & -- 17/9  &    2    &    10/9  & -- 14/9   \\
 4 & -- 2    &    38/9  &    8/9   &    --   &    23/9  &    11/9   \\
 5 & -- 2/3  &    10/9  &    4/9   & -- 2/3  &    4/9   &    4/9    \\
 6 & -- 4/3  & -- 29/9  & -- 11/9  & -- 4/3  & -- 11/9  & -- 11/9   \\
 7 &    11/9 &    35/54 & -- 13/54 &    11/9 & -- 31/54 & -- 13/54  \\
 8 & -- 4/9  &    22/27 &    22/27 & -- 4/9  &    34/27 &    22/27  \\
 9 &    --   &    --    &    --    &    --   &    --    &    --     \\
10 &    --   &    --    &    --    &    --   &    --    &    --     \\
\hline
\end{tabular}
\end{center}
\end{table}

\begin{table}[h]
\caption[]{
\label{tab:3}}
\begin{center}
\begin{tabular}{|c||c|c|c||c|c|c|}
\hline
&
\multicolumn{3}{c||}{$Q_{2} \longrightarrow V_{LR}$} &
\multicolumn{3}{c|}{$\widetilde{Q}_{6} \longrightarrow V_{LR}$} \\
\hline
$P$ &
$ 1/\eps^{2} $ &
$\left( 1/\eps \right)_{\rm NDR}$ &
$\left( 1/\eps \right)_{\rm HV}$  &
$ 1/\eps^{2} $ &
$\left( 1/\eps \right)_{\rm NDR}$ &
$\left( 1/\eps \right)_{\rm HV}$  \\
\hline
\hline
11 &    1  &    13/6 &   7/6  &    1  & -- 11/6 & -- 11/6  \\
12 & -- 1  &    5/6 &    11/6 & -- 1  & -- 7/6  & -- 7/6   \\
13 &    -- & -- 1   &    2/3  &    -- & -- 1    &    2/3   \\
14 &    2  & -- 2/3 & -- 1/3  &    2  & -- 8/3  & -- 1/3   \\
\hline
\end{tabular}
\end{center}
\end{table}

\begin{table}[h]
\caption[]{
\label{tab:4}}
\begin{center}
\begin{tabular}{|c||c|c|c||c|c|c|}
\hline
&
\multicolumn{3}{c||}{$Q_{2} \longrightarrow V_{LL}$} &
\multicolumn{3}{c|}{$\widetilde{Q}_{6} \longrightarrow V_{LL}$} \\
\hline
$P$ &
$ 1/\eps^{2} $ &
$\left( 1/\eps \right)_{\rm NDR}$ &
$\left( 1/\eps \right)_{\rm HV}$  &
$ 1/\eps^{2} $ &
$\left( 1/\eps \right)_{\rm NDR}$ &
$\left( 1/\eps \right)_{\rm HV}$  \\
\hline
\hline
11 &    1  & -- 5/6  & -- 11/6 &    1  &    7/6  &  7/6  \\
12 & -- 1  & -- 13/6 & -- 7/6  & -- 1  &    11/6 &  11/6 \\
13 & -- 2  & -- 4/3  &    1    & -- 2  &    2/3  &  1    \\
14 &    -- & -- 1    &    --   &    -- & -- 1    &  --   \\
\hline
\end{tabular}
\end{center}
\end{table}

\begin{table}[h]
\caption[]{Full QCD Anomalous Dimension Matrix $(\gss)_{ij}$
for the NDR and HV scheme (with vanishing entries omitted tacitly).
\label{tab:5}}
\begin{center}
\begin{tabular}{|c|c|c|}
\hline \tvs
$(i,j)$ & {\bf NDR} & {\bf HV} \tvs \\
\hline\hline \tvs
$(1,1)$ & $-{{22}\over 3} - {{57}\over {2 {N^2}}} - {{2 f}\over {3 N}}$ &\
  $-{{110}\over 3} - {{57}\over {2 {N^2}}} + {{44 {N^2}}\over 3} + \left(\
  {{14}\over {3 N}} - {{8 N}\over 3} \right)  f$ \tvs \\
$(1,2)$ & ${{39}\over N} - {{19 N}\over 6} + {{2 f}\over 3}$ & ${{39}\over N}\
  + {{23 N}\over 2} - 2 f$ \tvs \\
$(1,3)$ & ${{-2}\over {3 N}} + 3 N$ & ${{-4}\over N} + 3 N$ \tvs \\
$(1,4)$ & $-{7\over 3}$ & $1$ \tvs \\
$(1,5)$ & ${{16}\over {3 N}} - 3 N$ & ${2\over N} - 3 N$ \tvs \\
$(1,6)$ & $-{7\over 3}$ & $1$ \tvs \\
\hline \tvs
$(2,1)$ & ${{39}\over N} - {{19 N}\over 6} + {{2 f}\over 3}$ & ${{39}\over N}\
  + {{23 N}\over 2} - 2 f$ \tvs \\
$(2,2)$ & $-{{22}\over 3} - {{57}\over {2 {N^2}}} - {{2 f}\over {3 N}}$ &\
  $-{{110}\over 3} - {{57}\over {2 {N^2}}} + {{44 {N^2}}\over 3} + \left(\
  {{14}\over {3 N}} - {{8 N}\over 3} \right)  f$ \tvs \\
$(2,3)$ & $-{{32}\over {27}} + {{86}\over {27 {N^2}}}$ & $-{{56}\over {27}} +\
  {{86}\over {27 {N^2}}}$ \tvs \\
$(2,4)$ & ${{-230}\over {27 N}} + {{176 N}\over {27}}$ & ${{-140}\over {27\
  N}} + {{110 N}\over {27}}$ \tvs \\
$(2,5)$ & $-{{122}\over {27}} - {{94}\over {27 {N^2}}}$ & $-{{128}\over {27}}\
  - {{58}\over {27 {N^2}}}$ \tvs \\
$(2,6)$ & ${{130}\over {27 N}} + {{86 N}\over {27}}$ & ${{148}\over {27 N}} +\
  {{38 N}\over {27}}$ \tvs \\
\hline \tvs
$(3,3)$ & $-{{262}\over {27}} - {{1195}\over {54 {N^2}}} + \left( {{-10}\over\
  {3 N}} + 3 N \right)  f$ & $-{{1102}\over {27}} - {{1195}\over {54 {N^2}}}\
  + {{44 {N^2}}\over 3} + \left( {2\over {3 N}} + {N\over 3} \right)  f$ \tvs
\\
$(3,4)$ & ${{593}\over {27 N}} + {{533 N}\over {54}} + {f\over 3}$ &\
  ${{773}\over {27 N}} + {{1061 N}\over {54}} - f$ \tvs \\
$(3,5)$ & $-{{244}\over {27}} - {{188}\over {27 {N^2}}} + \left( {{10}\over\
  {3 N}} - 3 N \right)  f$ & $-{{256}\over {27}} - {{116}\over {27 {N^2}}} +\
  \left( {2\over N} - 3 N \right)  f$ \tvs \\
$(3,6)$ & ${{260}\over {27 N}} + {{172 N}\over {27}} - {f\over 3}$ &\
  ${{296}\over {27 N}} + {{76 N}\over {27}} + f$ \tvs \\
\hline \tvs
$(4,3)$ & ${{113}\over {3 N}} + {{17 N}\over 6} + \left( -{2\over {27}} +\
  {{74}\over {27 {N^2}}} \right)  f$ & ${{31}\over N} + {{35 N}\over 2} +\
  \left( -{{110}\over {27}} + {{86}\over {27 {N^2}}} \right)  f$ \tvs \\
$(4,4)$ & $-12 - {{57}\over {2 {N^2}}} + \left( {{-182}\over {27 N}} + {{110\
  N}\over {27}} \right)  f$ & $-{{104}\over 3} - {{57}\over {2 {N^2}}} + {{44\
  {N^2}}\over 3} + \left( {{-14}\over {27 N}} + {{38 N}\over {27}} \right) \
  f$ \tvs \\
$(4,5)$ & ${{32}\over {3 N}} - 6 N + \left( -{{56}\over {27}} + {2\over {27\
  {N^2}}} \right)  f$ & ${4\over N} - 6 N + \left( -{{128}\over {27}} -\
  {{58}\over {27 {N^2}}} \right)  f$ \tvs \\
$(4,6)$ & $-{{14}\over 3} + \left( {{-20}\over {27 N}} + {{74 N}\over {27}}\
  \right)  f$ & $2 + \left( {{148}\over {27 N}} + {{38 N}\over {27}} \right) \
  f$ \tvs \\
\hline \tvs
$(5,3)$ & $\left( {{20}\over {3 N}} - 3 N \right)  f$ & $\left( {8\over {3\
  N}} - 3 N \right)  f$ \tvs \\
$(5,4)$ & ${{-11 f}\over 3}$ & ${f\over 3}$ \tvs \\
$(5,5)$ & ${{137}\over 6} + {{15}\over {2 {N^2}}} + \left( {{-20}\over {3 N}}\
  + 3 N \right)  f$ & $-{{71}\over 6} + {{15}\over {2 {N^2}}} + {{44\
  {N^2}}\over 3} + {{N f}\over 3}$ \tvs \\
$(5,6)$ & ${3\over N} - {{100 N}\over 3} + {{11 f}\over 3}$ & ${3\over N} -\
  {{40 N}\over 3} - {f\over 3}$ \tvs \\
\hline \tvs
$(6,3)$ & $\left( -{{56}\over {27}} - {{178}\over {27 {N^2}}} \right)  f$ &\
  $\left( -{{128}\over {27}} - {{94}\over {27 {N^2}}} \right)  f$ \tvs \\
$(6,4)$ & $\left( {{250}\over {27 N}} - {{16 N}\over {27}} \right)  f$ &\
  $\left( {{202}\over {27 N}} + {{20 N}\over {27}} \right)  f$ \tvs \\
$(6,5)$ & ${{-18}\over N} - {{71 N}\over 2} + \left( {{178}\over {27}} +\
  {{74}\over {27 {N^2}}} \right)  f$ & ${{-18}\over N} + {{107 N}\over 6} +\
  \left( -{{74}\over {27}} + {{86}\over {27 {N^2}}} \right)  f$ \tvs \\
$(6,6)$ & ${{479}\over 6} + {{15}\over {2 {N^2}}} - {{203 {N^2}}\over 6} +\
  \left( {{-452}\over {27 N}} + {{200 N}\over {27}} \right)  f$ &\
  $-{{17}\over 6} + {{15}\over {2 {N^2}}} - {{9 {N^2}}\over 2} + \left(\
  {{-68}\over {27 N}} + {{56 N}\over {27}} \right)  f$ \tvs \\
\hline
\end{tabular}
\end{center}
\end{table}

\addtocounter{table}{-1}
\begin{table}[h]
\caption[]{Full QCD Anomalous Dimension Matrix $(\gss)_{ij}$
for the NDR and HV scheme (continued; with vanishing entries omitted
tacitly).
}
\begin{center}
\small
\hskip -1.7cm
\begin{tabular}{|c|c|c|}
\hline \tvs
$(i,j)$ & {\bf NDR} & {\bf HV} \tvs \\
\hline\hline \tvs
$(7,3)$ & $\left( {{-10}\over {3 N}} + {{3 N}\over 2} \right)  d + \left(\
  {{20}\over {3 N}} - 3 N \right)  u$ & $\left( {{-4}\over {3 N}} + {{3\
  N}\over 2} \right)  d + \left( {8\over {3 N}} - 3 N \right)  u$ \tvs \\
$(7,4)$ & ${{11 d}\over 6} - {{11 u}\over 3}$ & ${{-d}\over 6} + {u\over 3}$\
  \tvs \\
$(7,5)$ & $\left( {{-1}\over {3 N}} - {{3 N}\over 2} \right)  d + \left(\
  {2\over {3 N}} + 3 N \right)  u$ & $\left( {5\over {3 N}} - {{3 N}\over 2}\
  \right)  d + \left( {{-10}\over {3 N}} + 3 N \right)  u$ \tvs \\
$(7,6)$ & ${{11 d}\over 6} - {{11 u}\over 3}$ & ${{-d}\over 6} + {u\over 3}$\
  \tvs \\
$(7,7)$ & ${{137}\over 6} + {{15}\over {2 {N^2}}} - {{22 f}\over {3 N}}$ &\
  $-{{71}\over 6} + {{15}\over {2 {N^2}}} + {{44 {N^2}}\over 3} + \left(\
  {{10}\over {3 N}} - {{8 N}\over 3} \right)  f$ \tvs \\
$(7,8)$ & ${3\over N} - {{100 N}\over 3} + {{22 f}\over 3}$ & ${3\over N} -\
  {{40 N}\over 3} - {{2 f}\over 3}$ \tvs \\
\hline \tvs
$(8,3)$ & $\left( {{28}\over {27}} + {{89}\over {27 {N^2}}} \right)  d +\
  \left( -{{56}\over {27}} - {{178}\over {27 {N^2}}} \right)  u$ & $\left(\
  {{64}\over {27}} + {{47}\over {27 {N^2}}} \right)  d + \left( -{{128}\over\
  {27}} - {{94}\over {27 {N^2}}} \right)  u$ \tvs \\
$(8,4)$ & $\left( {{-125}\over {27 N}} + {{8 N}\over {27}} \right)  d +\
  \left( {{250}\over {27 N}} - {{16 N}\over {27}} \right)  u$ & $\left(\
  {{-101}\over {27 N}} - {{10 N}\over {27}} \right)  d + \left( {{202}\over\
  {27 N}} + {{20 N}\over {27}} \right)  u$ \tvs \\
$(8,5)$ & $\left( -{{35}\over {27}} - {{37}\over {27 {N^2}}} \right)  d +\
  \left( {{70}\over {27}} + {{74}\over {27 {N^2}}} \right)  u$ & $\left(\
  {{19}\over {27}} - {{43}\over {27 {N^2}}} \right)  d + \left( -{{38}\over\
  {27}} + {{86}\over {27 {N^2}}} \right)  u$ \tvs \\
$(8,6)$ & $\left( {{127}\over {27 N}} - {{55 N}\over {27}} \right)  d +\
  \left( {{-254}\over {27 N}} + {{110 N}\over {27}} \right)  u$ & $\left(\
  {{79}\over {27 N}} - {{55 N}\over {27}} \right)  d + \left( {{-158}\over\
  {27 N}} + {{110 N}\over {27}} \right)  u$ \tvs \\
$(8,7)$ & ${{-18}\over N} - {{71 N}\over 2} + 4 f$ & ${{-18}\over N} + {{107\
  N}\over 6} - {{4 f}\over 3}$ \tvs \\
$(8,8)$ & ${{479}\over 6} + {{15}\over {2 {N^2}}} - {{203 {N^2}}\over 6} +\
  \left( {{-22}\over {3 N}} + {{10 N}\over 3} \right)  f$ & $-{{17}\over 6} +\
  {{15}\over {2 {N^2}}} - {{9 {N^2}}\over 2} + \left( {{10}\over {3 N}} - 2 N\
  \right)  f$ \tvs \\
\hline \tvs
$(9,3)$ & ${{32}\over {27}} - {{86}\over {27 {N^2}}} + \left( {4\over {3 N}}\
  - {{3 N}\over 2} \right)  d + \left( {{-8}\over {3 N}} + 3 N \right)  u$ &\
  ${{56}\over {27}} - {{86}\over {27 {N^2}}} + \left( {2\over N} - {{3\
  N}\over 2} \right)  d + \left( {{-4}\over N} + 3 N \right)  u$ \tvs \\
$(9,4)$ & ${{230}\over {27 N}} - {{176 N}\over {27}} + {d\over 6} - {u\over\
  3}$ & ${{140}\over {27 N}} - {{110 N}\over {27}} - {d\over 2} + u$ \tvs \\
$(9,5)$ & ${{122}\over {27}} + {{94}\over {27 {N^2}}} + \left( {{-5}\over {3\
  N}} + {{3 N}\over 2} \right)  d + \left( {{10}\over {3 N}} - 3 N \right)
  u$ & ${{128}\over {27}} + {{58}\over {27 {N^2}}} + \left( -{1\over N} + {{3\
  N}\over 2} \right)  d + \left( {2\over N} - 3 N \right)  u$ \tvs \\
$(9,6)$ & ${{-130}\over {27 N}} - {{86 N}\over {27}} + {d\over 6} - {u\over\
  3}$ & ${{-148}\over {27 N}} - {{38 N}\over {27}} - {d\over 2} + u$ \tvs \\
$(9,9)$ & $-{{22}\over 3} - {{57}\over {2 {N^2}}} - {{2 f}\over {3 N}}$ &\
  $-{{110}\over 3} - {{57}\over {2 {N^2}}} + {{44 {N^2}}\over 3} + \left(\
  {{14}\over {3 N}} - {{8 N}\over 3} \right)  f$ \tvs \\
$(9,10)$ & ${{39}\over N} - {{19 N}\over 6} + {{2 f}\over 3}$ & ${{39}\over\
  N} + {{23 N}\over 2} - 2 f$ \tvs \\
\hline \tvs
$(10,3)$ & ${2\over {3 N}} - 3 N + \left( {{10}\over {27}} - {{37}\over {27\
  {N^2}}} \right)  d + \left( -{{20}\over {27}} + {{74}\over {27 {N^2}}}\
  \right)  u$ & ${4\over N} - 3 N + \left( {{28}\over {27}} - {{43}\over {27\
  {N^2}}} \right)  d + \left( -{{56}\over {27}} + {{86}\over {27 {N^2}}}\
  \right)  u$ \tvs \\
$(10,4)$ & ${7\over 3} + \left( {{82}\over {27 N}} - {{55 N}\over {27}}\
  \right)  d + \left( {{-164}\over {27 N}} + {{110 N}\over {27}} \right)  u$\
  & $-1 + \left( {{70}\over {27 N}} - {{55 N}\over {27}} \right)  d + \left(\
  {{-140}\over {27 N}} + {{110 N}\over {27}} \right)  u$ \tvs \\
$(10,5)$ & ${{-16}\over {3 N}} + 3 N + \left( {{28}\over {27}} - {1\over {27\
  {N^2}}} \right)  d + \left( -{{56}\over {27}} + {2\over {27 {N^2}}} \right)
   u$ & ${{-2}\over N} + 3 N + \left( {{64}\over {27}} + {{29}\over {27\
  {N^2}}} \right)  d + \left( -{{128}\over {27}} - {{58}\over {27 {N^2}}}\
  \right)  u$ \tvs \\
$(10,6)$ & ${7\over 3} + \left( {{10}\over {27 N}} - {{37 N}\over {27}}\
  \right)  d + \left( {{-20}\over {27 N}} + {{74 N}\over {27}} \right)  u$ &\
  $-1 + \left( {{-74}\over {27 N}} - {{19 N}\over {27}} \right)  d + \left(\
  {{148}\over {27 N}} + {{38 N}\over {27}} \right)  u$ \tvs \\
$(10,9)$ & ${{39}\over N} - {{19 N}\over 6} + {{2 f}\over 3}$ & ${{39}\over\
  N} + {{23 N}\over 2} - 2 f$ \tvs \\
$(10,10)$ & $-{{22}\over 3} - {{57}\over {2 {N^2}}} - {{2 f}\over {3 N}}$ &\
  $-{{110}\over 3} - {{57}\over {2 {N^2}}} + {{44 {N^2}}\over 3} + \left(\
  {{14}\over {3 N}} - {{8 N}\over 3} \right)  f$ \tvs \\
\hline
\end{tabular}
\end{center}
\end{table}

\clearpage
\newsection{Two--Loop QCD Anomalous Dimension Matrix $\gss$ in NDR and
HV Schemes for $N=3$}
\begin{displaymath}
\gssndr\bigl|_{N=3} =
\left(
\begin{array}{ccccc}
-{{21}\over 2} - {{2\,f}\over 9} & {7\over 2} + {{2\,f}\over 3} & {{79}\over\
  9} & -{7\over 3} & -{{65}\over 9} \\ \mvs
{7\over 2} + {{2\,f}\over 3} & -{{21}\over 2} - {{2\,f}\over 9} &\
  -{{202}\over {243}} & {{1354}\over {81}} & -{{1192}\over {243}} \\ \mvs
0 & 0 & -{{5911}\over {486}} + {{71\,f}\over 9} & {{5983}\over {162}} +\
  {f\over 3} & -{{2384}\over {243}} - {{71\,f}\over 9} \\ \mvs
0 & 0 & {{379}\over {18}} + {{56\,f}\over {243}} & -{{91}\over 6} +\
  {{808\,f}\over {81}} & -{{130}\over 9} - {{502\,f}\over {243}} \\ \mvs
0 & 0 & {{-61\,f}\over 9} & {{-11\,f}\over 3} & {{71}\over 3} + {{61\,f}\over\
  9} \\ \mvs
0 & 0 & {{-682\,f}\over {243}} & {{106\,f}\over {81}} & -{{225}\over 2} +\
  {{1676\,f}\over {243}} \\ \mvs
0 & 0 & {{-61\,(u-d/2)}\over 9} & {{-11\,(u-d/2)}\over 3} & {{83\,
(u-d/2)}\over 9} \\ \mvs
0 & 0 & {{-682\,(u-d/2)}\over {243}} & {{106\,(u-d/2)}\over {81}} &\
  {{704\,(u-d/2)}\over {243}} \\ \mvs
0 & 0 & {{202}\over {243}} + {{73\,(u-d/2)}\over 9} & -{{1354}\over {81}} -\
  {{(u-d/2)}\over 3} & {{1192}\over {243}} - {{71\,(u-d/2)}\over 9} \\ \mvs
0 & 0 & -{{79}\over 9} - {{106\,(u-d/2)}\over {243}} & {7\over 3} +\
  {{826\,(u-d/2)}\over {81}} & {{65}\over 9} - {{502\,(u-d/2)}\over\
  {243}}
\end{array}
\right.
\end{displaymath}

\bigskip

\begin{displaymath}
\phantom{\gssndr\bigl|_{N=3} =}
\left.
\begin{array}{ccccc}
-{7\over 3} & 0 & 0 & 0 & 0 \\ \mvs
{{904}\over {81}} & 0 & 0 & 0 & 0 \\ \mvs
{{1808}\over {81}} - {f\over 3} & 0 & 0 & 0 & 0 \\ \mvs
-{{14}\over 3} + {{646\,f}\over {81}} & 0 & 0 & 0 & 0 \\ \mvs
-99 + {{11\,f}\over 3} & 0 & 0 & 0 & 0 \\ \mvs
-{{1343}\over 6} + {{1348\,f}\over {81}} & 0 & 0 & 0 & 0 \\ \mvs
{{-11\,(u-d/2)}\over 3} & {{71}\over 3} - {{22\,f}\over 9} & -99 +\
  {{22\,f}\over 3} & 0 & 0 \\ \mvs
{{736\,(u-d/2)}\over {81}} & -{{225}\over 2} + 4\,f & -{{1343}\over 6} +\
  {{68\,f}\over 9} & 0 & 0 \\ \mvs
-{{904}\over {81}} - {{(u-d/2)}\over 3} & 0 & 0 & -{{21}\over 2} -\
  {{2\,f}\over 9} & {7\over 2} + {{2\,f}\over 3} \\ \mvs
{7\over 3} + {{646\,(u-d/2)}\over {81}} & 0 & 0 & {7\over 2} + {{2\,f}\over\
  3} & -{{21}\over 2} - {{2\,f}\over 9}
\end{array}
\right)
\end{displaymath}
\newpage
\begin{displaymath}
\gsshv\bigl|_{N=3} =
\left(
\begin{array}{ccccc}
{{553}\over 6} - {{58\,f}\over 9} & {{95}\over 2} - 2\,f & {{23}\over 3} & 1\
  & -{{25}\over 3} \\ \mvs
{{95}\over 2} - 2\,f & {{553}\over 6} - {{58\,f}\over 9} & -{{418}\over\
  {243}} & {{850}\over {81}} & -{{1210}\over {243}} \\ \mvs
0 & 0 & {{43121}\over {486}} + {{11\,f}\over 9} & {{11095}\over {162}} - f &\
  -{{2420}\over {243}} - {{25\,f}\over 3} \\ \mvs
0 & 0 & {{377}\over 6} - {{904\,f}\over {243}} & {{565}\over 6} +\
  {{328\,f}\over {81}} & -{{50}\over 3} - {{1210\,f}\over {243}} \\ \mvs
0 & 0 & {{-73\,f}\over 9} & {f\over 3} & 121 + f \\ \mvs
0 & 0 & {{-1246\,f}\over {243}} & {{382\,f}\over {81}} & {{95}\over 2} -\
  {{580\,f}\over {243}} \\ \mvs
0 & 0 & {{-73\,(u-d/2)}\over 9} & {{(u-d/2)}\over 3} & {{71\,(u-d/2)}\over 9}
\\ \mvs
0 & 0 & {{-1246\,(u-d/2)}\over {243}} & {{382\,(u-d/2)}\over {81}} &\
  {{-256\,(u-d/2)}\over {243}} \\ \mvs
0 & 0 & {{418}\over {243}} + {{23\,(u-d/2)}\over 3} & -{{850}\over {81}} +\
  (u-d/2) & {{1210}\over {243}} - {{25\,(u-d/2)}\over 3} \\ \mvs
0 & 0 & -{{23}\over 3} - {{418\,(u-d/2)}\over {243}} & -1 + {{850\,(u-d/2)}
\over {81}} & {{25}\over 3} - {{1210\,(u-d/2)}\over {243}}
\end{array}
\right.
\end{displaymath}

\bigskip

\begin{displaymath}
\phantom{\gsshv\bigl|_{N=3} =}
\left.
\begin{array}{ccccc}
1 & 0 & 0 & 0 & 0 \\ \mvs
{{490}\over {81}} & 0 & 0 & 0 & 0 \\ \mvs
{{980}\over {81}} + f & 0 & 0 & 0 & 0 \\ \mvs
2 + {{490\,f}\over {81}} & 0 & 0 & 0 & 0 \\ \mvs
-39 - {f\over 3} & 0 & 0 & 0 & 0 \\ \mvs
-{{85}\over 2} + {{436\,f}\over {81}} & 0 & 0 & 0 & 0 \\ \mvs
{{(u-d/2)}\over 3} & 121 - {{62\,f}\over 9} & -39 - {{2\,f}\over 3} & 0 & 0\
  \\ \mvs
{{832\,(u-d/2)}\over {81}} & {{95}\over 2} - {{4\,f}\over 3} & -{{85}\over\
  2} - {{44\,f}\over 9} & 0 & 0 \\ \mvs
-{{490}\over {81}} + (u-d/2) & 0 & 0 & {{553}\over 6} - {{58\,f}\over 9} &\
  {{95}\over 2} - 2\,f \\ \mvs
-1 + {{490\,(u-d/2)}\over {81}} & 0 & 0 & {{95}\over 2} - 2\,f &\
  {{553}\over 6} - {{58\,f}\over 9}
\end{array}
\right)
\end{displaymath}

\clearpage
\newsection{Figures of Feynman Diagrams}
\begin{figure}[h]
\vspace{0.15in}
\centerline{
\epsfysize=1.7in
\epsffile{fig1qcd.ps}
}
\vspace{0.15in}
\caption[]{
\label{fig:1}}
\end{figure}

\begin{figure}[h]
\vspace{0.15in}
\centerline{
\epsfysize=1.7in 
\epsffile{fig2qcd.ps}
}
\vspace{0.15in}
\caption[]{
\label{fig:2}}
\end{figure}

\begin{figure}[h]
\vspace{0.15in}
\centerline{
\epsfysize=1.7in 
\epsffile{fig3qcd.ps}
}
\vspace{0.15in}
\caption[]{
\label{fig:3}}
\end{figure}

\newpage

\begin{figure}[h]
\vspace{0.15in}
\centerline{
\epsfysize=8in
\epsffile{fig4qcd.ps}
}
\vspace{0.15in}
\caption[]{
\label{fig:4}}
\end{figure}

\newpage

\begin{figure}[h]
\vspace{0.15in}
\centerline{
\epsfysize=8in
\epsffile{fig5qcd.ps}
}
\vspace{0.15in}
\caption[]{
\label{fig:5}}
\end{figure}

\newpage

\begin{figure}[h]
\vspace{0.15in}
\centerline{
\epsfysize=1.7in 
\epsffile{fig6qcd.ps}
}
\vspace{0.15in}
\caption[]{
\label{fig:6}}
\end{figure}

\begin{figure}[h]
\vspace{0.15in}
\centerline{
\epsfysize=5.1in 
\epsffile{fig7qcd.ps}
}
\vspace{0.15in}
\caption[]{
\label{fig:7}}
\end{figure}

\clearpage
\centerline{\Large\bf Figure Captions}

\bigskip

\begin{description}
\item[Figure 1:]
The three basic ways of inserting a given operator into a four--point
function: (a) current--current-, (b) type 1 penguin-, (c) type 2
penguin-insertion.  The curled lines denote gluons. The 4-vertices
``$\otimes\ \otimes$'' denote standard operator insertions.
\item[Figure 2:]
One--loop current--current diagrams contributing to $\gs$.
The meaning of lines and vertices is the same as in fig.~1.
Possible left-light or up-down reflected diagrams are not shown.
\item[Figure 3:]
One--loop type 1 and 2 penguin diagrams contributing to $\gs$.
The meaning of lines and vertices is the same as in fig.~1.
\item[Figure 4:]
Two--loop current--current diagrams contributing to $\gss$.
The meaning of lines and vertices is the same as in fig.~1.
In addition shaded blobs stand for self-energy insertions.
Possible left-light or up-down reflected diagrams are not shown.
\item[Figure 5:]
Two--loop penguin diagrams contributing to $\gss$.
The curled lines denote gluons. Square-vertices stand for type 1 and 2
penguin insertions as of figs.~\ref{fig:1}(b) and (c), respectively.
Possible left-light reflected diagrams are not shown.
\item[Figure 6:]
Two--loop penguin diagrams having no $1/\eps$ divergence and hence do
not contribute to the anomalous dimension matrix $\gss$.  The meaning of
lines and vertices is the same as in fig.~5.  In addition shaded blobs
stand for self-energy insertions or vertex corrections.
\item[Figure 7:]
Two--loop penguin diagrams vanishing identically in dimensional
regularization.  The meaning of lines and vertices is the same as in
fig.~5. In addition shaded blobs stand for self-energy insertions or
vertex corrections.
\end{description}

\end{document}